\documentclass[%
 reprint,
superscriptaddress,
aip,
jap,
]{revtex4-2}
\usepackage{graphicx}
\usepackage{dcolumn}
\usepackage{bm}
\usepackage{booktabs}
\usepackage{xcolor}
\usepackage{siunitx}
\usepackage[utf8]{inputenc}
\usepackage[T1]{fontenc}
\usepackage{mathptmx}
\usepackage{physics}
\usepackage{mathtools}
\usepackage{bm}
\usepackage[normalem]{ulem}

\begin{document}

\preprint{APS/123-QED}

\title{Artificial Spin Ice: A Tutorial on Design and Control of Geometry, Microstate, Magnon Dynamics \& Neuromorphic Computing}

\author{Rawnak Sultana}
\affiliation{Department of Physics and Astronomy, University of Delaware, Newark, DE 19716, USA}%
\author{Amrit Kumar Mondal}
\affiliation{Department of Physics and Astronomy, University of Delaware, Newark, DE 19716, USA}%
\author{Vinayak Shantaram Bhat}%
\affiliation{Department of Physics and Astronomy, University of Delaware, Newark, DE 19716, USA}%
\author{Kilian Stenning}
\affiliation{Blackett Laboratory, Imperial College London, London, UK}%
\affiliation{London Centre for Nanotechnology, Imperial College London, London, United Kingdom}
\author{Yue Li}
\affiliation{Materials Science Division, Argonne National Laboratory, Lemont, IL 60439, USA}
\author{Daan M. Arroo}
\affiliation{
Department of Materials, Imperial College London, Exhibition Road, London SW7 2AZ, United Kingdom}
\affiliation{London Centre for Nanotechnology, Imperial College London, London, United Kingdom}
\author{Aastha Vasdev}
\affiliation{Materials Science Division, Argonne National Laboratory, Lemont, IL 60439, USA}
\affiliation{Department of Electrical and Computer Engineering, University of Kentucky, Lexington, KY 40506, USA}
\author{Margaret R. McCarter}
\affiliation{Department of Physics and Astronomy, University of Kentucky, Lexington, KY 40506, USA}
\affiliation{Advanced Light Source, Lawrence Berkeley National Laboratory, Berkeley, CA 94720, USA}
\author{Lance E. De Long}%
\affiliation{Department of Physics and Astronomy, University of Kentucky, Lexington, KY 40506, USA}%
\author{J. Todd Hastings}%
\affiliation{Department of Physics and Astronomy, University of Kentucky, Lexington, KY 40506, USA}%
\author{Jack C. Gartside}
\affiliation{Blackett Laboratory, Imperial College London, London, UK}
\affiliation{London Centre for Nanotechnology, Imperial College London, London, United Kingdom}
\affiliation{Institute for Materials Research, Tohoku University, Sendai, Japan}
\author{M. Benjamin Jungfleisch}
\email{mbj@udel.edu}
\affiliation{Department of Physics and Astronomy, University of Delaware, Newark, DE 19716, USA}%

\date{\today}

\begin{abstract}
Artificial spin ice, arrays of strongly interacting nanomagnets, are complex magnetic systems with many emergent properties, rich microstate spaces, intrinsic physical memory, high-frequency dynamics in the GHz range and compatibility with a broad range of measurement approaches. This tutorial article aims to provide the foundational knowledge needed to understand, design, develop, and improve the dynamic properties of artificial spin ice (ASI). Special emphasis is placed on introducing the theory of micromagnetics, which describes the complex dynamics within these systems, along with their design, fabrication methods, and standard measurement and control techniques. The article begins with a review of the historical background, introducing the underlying physical phenomena and interactions that govern artificial spin ice. We then explore standard experimental techniques used to prepare the microstate space of the nanomagnetic array and to characterize magnetization dynamics, both in artificial spin ice and more broadly in ferromagnetic materials. Finally, we introduce the basics of neuromorphic computing applied to the case of artificial spin ice systems with goal to help researchers new to the field grasp these exciting new developments.

\end{abstract}

\maketitle


\section{\label{sec:intro}Introduction}\label{intro}
Artificial spin ice (ASI) \cite{gliga2020dynamics, sklenar2019dynamics, skjaervo2020advances, zhou2016large, wang2007} systems are arrays of lithographically fabricated two- (2D) or three-dimensional (3D) nanomagnets arranged in periodic or aperiodic lattices, providing a unique platform to study exotic magnetic states not found in natural materials. Initially conceived as a mesoscopic analog to magnetically frustrated rare-earth pyrochlores \cite{harris1997geometrical, bramwell2001spin} - crystalline counterparts to water ice \cite{slater2019surface} -- ASI has since evolved into a rich and independent field of study. These arrays are typically modeled as interacting binary Ising-like macrospins within nanoislands, or vortex states. ASI enables the exploration of phenomena such as geometrical frustration, emergent magnetic monopoles and antimonopoles connected by Dirac strings, and phase transitions. These systems also show immense promise as reprogrammable magnonic crystals, where the coherent and collective excitation of spin waves (SWs) serves as information carriers \cite{rana2021applications, lendinez2019magnetization, gubbiotti20242025, barman20212021}. 

\begin{figure}[h]
	\includegraphics[width=0.5\textwidth]{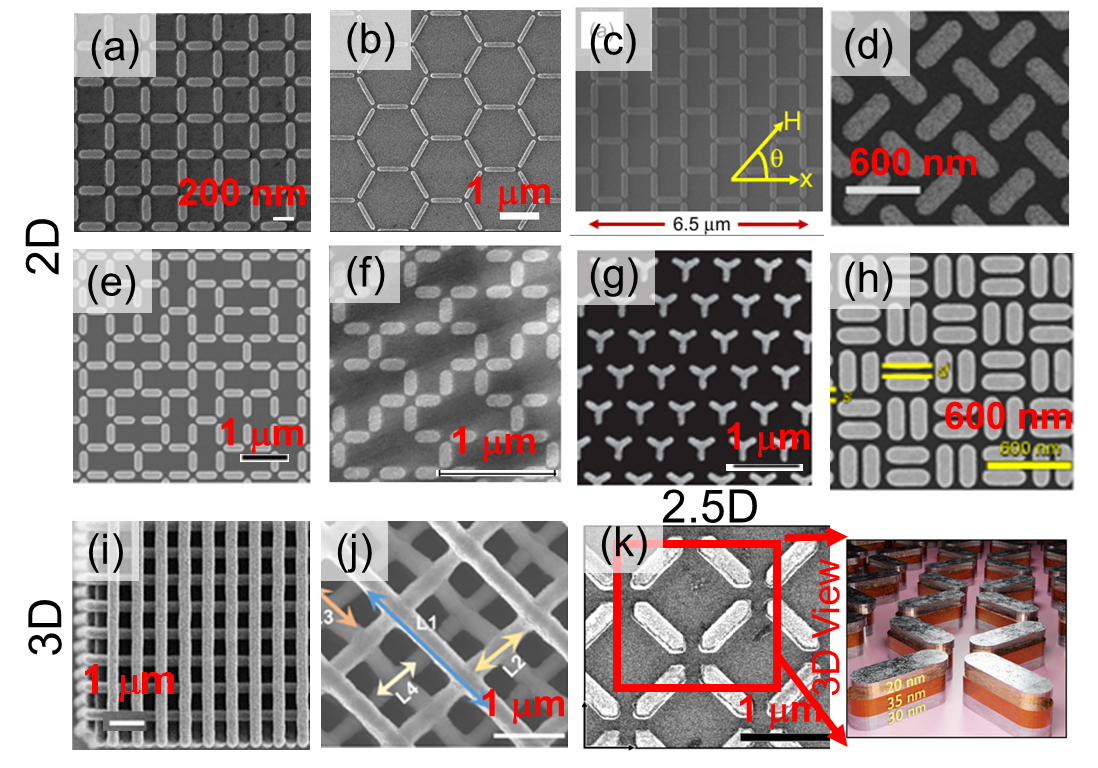}
	\centering
	\caption{Various types of artificial spin ice arranged on an (a) square, (b) Kagome, (c) brickwork, (d) pinwheel, (e) Santa Fe, (f) Tetris, (g) tripod, (h) ferroquadrupolar lattice, (i) polymeric nano-scaffold, (j) diamond-lattice, and (k) trilayer square lattice. The SEM images, starting from brickwork to the trilayer square lattice are adapted from Refs.~[\onlinecite{Park_2017, gliga2017emergent, zhang2021string, goryca2023deconstructing, zhang2023artificial, sklenar2019field, guo2023realization, sahoo2021observation, dion2024ultrastrong}]. Typical lateral dimensions of individual elements are $300\times 80$~nm$^2$. }\label{Figure_1}
\end{figure}

Unlike electromagnetic (EM) waves, SWs exhibit much slower propagation velocities and significantly shorter wavelengths (on the micro- to nanometer scale) at the same frequency, making them an ideal fit for nanotechnology. This compatibility opens pathways for on-chip data transfer and GHz frequency information processing while reducing power dissipation. Given these advantages, the study of magnetization dynamics in various 2D- and 2.5D- (multilayer 2D systems, defined as 2.5D in this tutorial) ASI geometries -- such as square, kagome, shakti, Tetris, Santa Fe, brickwork lattices and so on -- as well as in 3D ASI systems, have gained increasing attention (Fig.~\ref{Figure_1}) \cite{gartside2022reconfigurable, mondal2024brillouin, lendinez2023nonlinear, bhat2016magnetization, bang2020influence, Park_2017, gliga2017emergent, mondal2024string, zhang2021string, goryca2023deconstructing, zhang2023artificial, sklenar2019field, guo2023realization, sahoo2021observation, dion2024ultrastrong}. These dynamics are influenced by factors like thermal fluctuations, external magnetic fields, and the material properties of the spin-ice lattice. These studies offer insights into both exotic fundamental physics and applied functionality, positioning ASI systems as promising candidates for reconfigurable magnonic circuits and neuromorphic computing platforms due to their non-linearity and fading memory characteristics \cite{stenning2024neuromorphic}.

\begin{figure}[t]
	\includegraphics[width=0.5\textwidth]{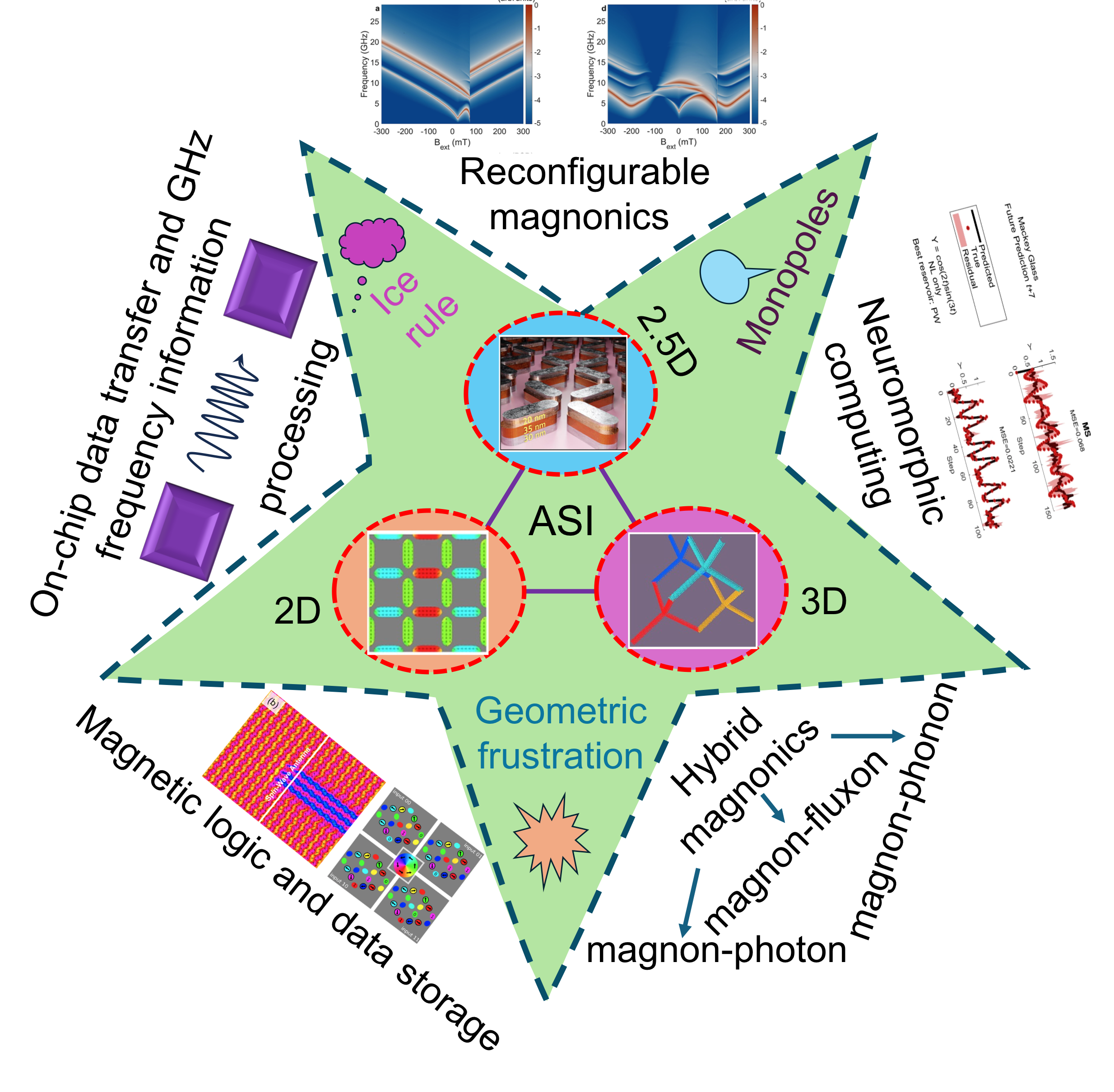}
	\centering
	\caption{Overview of research fields and potential applications in artificial-spin-ice based magentization dynamics. Adapted from Refs.~[\onlinecite{lendinez2023nonlinear, stenning2024neuromorphic, gypens2018balanced,li2022writable, mondal2024brillouin, dion2024ultrastrong, sahoo2021observation}].}.\label{Figure_2}
\end{figure}

In this tutorial article, we explicitly discuss the magnetization dynamics in ASI and how their rich magnetization states make them a promising candidate for realizing functional magnonic devices. Figure~\ref{Figure_2} illustrates the various branches of ASI research based on magnetization dynamics, focusing on both fundamental physics and potential applications. This tutorial article is organized into nine sections. Sections I, II, and III introduce the concept of ASI, its theoretical background, and the micromagnetic approach used to study magnetization dynamics in ASI. Sections IV and V cover key nanofabrication techniques for fabricating ASI samples and controlling their microstates. Sections VI, VII, and VIII delve into various static and dynamic measurement methods, supported by recent experimental findings on magnetization dynamics in ASIs, spanning from 2D to 3D systems, and extending to neuromorphic computing applications in ASI. Finally, Section IX offers a summary of the article and discusses potential future directions in the field.

\section{Theoretical Background}\label{theory}
The magnetic configurations of ferromagnetic nanoislands in ASI and their collective dynamics are governed by the Landau-Lifshitz-Gilbert (LLG) equation\cite{gliga2020dynamics,sklenar2019dynamics},

\begin{equation}
\frac{d\vec{M}}{dt} = -\abs{\gamma} (\vec{M} \cross \vec{H}_{eff})+\frac{\alpha}{ \ M_s} \left(\vec{M} \cross\frac{d\vec{M}}{dt}\right)\label{LLG}
\end{equation}

Here, the first term represents the precessional term, while the second term represents the damping term. In this context, $\gamma$ is the electron gyromagnetic ratio, 
$M$ is the magnetization, $M_s$ is the saturation magnetization, and $\alpha$ is a phenomenological dimensionless magnetic damping parameter. This form assumes that $\alpha\ll 1$, as is typically the case for systems of practical interest. Typically, the effective field contributions used in ASIs are\cite{gliga2020dynamics},
\begin{equation}
    \vec{H}_{eff} = \vec{H}_{Ex} + \vec{H}_{Ani} + \vec{H}_{Bias} + \vec{H}_{D}, \label{field}
\end{equation}    
which includes exchange ($H_{Ex}$), intrinsic anisotropy originating from crystalline spin-orbit coupling or from material structures such as layering, interfaces, or grain structures ($H_{Ani}$), an applied external field ($H_{Bias}$), and nonlocal magnetostatic (e.g., dipolar) fields ($H_{D}$).
The LLG equation, Eq.~(\ref{LLG}), subject to the effective field equation, Eq.~(\ref{field}), is generally a system of coupled nonlinear partial differential equations. Analytical solutions are typically found in cases
where the magnetostatic field is simplified, e.g., in the thin film limit where it reduces to a local field\cite{prabhakar2009spin}.  Consequently, numerical techniques are often required in order to solve the LLG equation. There are semi-analytical models\cite{Iacocca_2016,jungfleisch2016dynamic} and numerical tools like micromagnetic simulations for solving the LLG equation in ASI structure\cite{gliga2020dynamics}, which we discuss in the following.

\section{Micromagnetics}
The investigation of magnetization dynamics of ASIs through micromagnetic simulations has unveiled a significant revelation: the existence of effective magnetic monopoles within ASIs induces notable alterations in the ferromagnetic resonance spectrum \cite{gliga2013spectral}. Subsequent experimental analyses have not only corroborated these findings but also identified distinctive magnodynamic signatures, offering insights into the underlying physics of ASIs and their potential applications\cite{jungfleisch2016dynamic,bhat2016magnetization}. This underscores the indispensable role of micromagnetic simulations as a vital tool for comprehending and forecasting novel phenomena within ASIs, thereby advancing our understanding and exploration of these intriguing materials\cite{gliga2020dynamics}. Therefore, in the following, we present a brief review of micromagnetic simulation techniques to simulate magnetization dynamics in ASIs.


Classical electromagnetic theory postulates that the spontaneous magnetization is an intrinsic property of ferromagnetic (FM) materials, which imparts a magnetic moment to even the minutest volume of a material.
Initially approached in a somewhat empirical fashion, the concept of spontaneous magnetization found its first comprehensive explanation through the pioneering work of Weiss, who introduced the molecular field postulate. This postulate was later refined by Heisenberg, who, employing the principles of quantum mechanics, replaced the notion of the \textit{molecular field} with the concept of exchange forces. According to Heisenberg's theory, these exchange forces strive to align the spins of magnetic entities, counteracted by the disruptive influence of thermal energy, which tends to misalign them. Nonetheless, this theory lacks explicit directionality regarding magnetization as it only relies on neighboring spins being parallel \cite{brown1963micromagnetics}. 

Observations under constant temperature, viewed through microscopes, reveal a curious phenomenon: while magnetization remains uniform, its orientation varies across different regions, unless either a magnetic field is applied or the material's size falls below a critical threshold. This tendency for ferromagnetic thin films to organize into domains, each with uniform magnetization but differing orientations, points towards the formation of a demagnetized state, as initially postulated by Weiss. The domain theory advanced by Weiss effectively rationalizes such observations, yet it becomes apparent that the presence and dimensions of domains are significantly influenced by anisotropy forces within magnetic materials \cite{brown1963micromagnetics}.

Therefore, the pursuit of a comprehensive theory is called for, which is capable of bridging the intricate balance between atomic-scale exchange forces and macroscale dipolar forces.
Take, for instance, the behavior of thin, infinitely long ferromagnets, where spontaneous magnetization aligns parallel to the elongated axis, resulting in a single domain. Contrastingly, finite-sized ferromagnets provoke the emergence of surface poles along their boundaries, giving rise to magnetostatic energy ($E_{Ms}$). To mitigate $E_{Ms}$, alterations in spin distributions become imperative, consequently reshaping the single-domain behavior, and inducing an increase in exchange energy ($E_{Ex}$). The system reaches equilibrium when both exchange and magnetostatic energies are minimized. This quest for such explanation finds its answer in the phenomenological theory of micromagnetics, pioneered by Brown, which adeptly bridges these two length scales in a compelling manner.
 The aim of micromagnetics is to develop a formalism in which the macroscopic properties of a material can be simulated by introducing adequate approximations to the fundamental atomic behavior of the material  \cite{buschow2005concise}.  This, combined with an energy minimization method, forms a basis for classical micromagnetics  \cite{chikazumi2009physics}. 
\\

Micromagnetics incorporates two key conventions:
\begin{enumerate}
	\item The magnetization is treated as a continuous vector field that varies with position, expressed as $\overrightarrow{M}\equiv \overrightarrow{M}(x,y,z)$. 
	\item  The magnetization vector is normalized to have a unit length, satisfying the condition $M^{2}=1$.   
\end{enumerate}

\subsection{Energies and Magnetic Fields}
In micromagnetics, the equilibrium configuration of magnetization under the influence of various forces is determined using energy minimization principles. The free energy of a ferromagnetic body has four key contributions: exchange energy, demagnetization energy, anisotropy energy, and Zeeman energy (arising from an external magnetic field). The relationship between the total effective magnetic field and the corresponding energy is expressed as:
\begin{equation} \label{MM:001}
	E_{Tot} = -\int \mu_{o} \vec{M}\cdot \vec{H}_{eff} dV
\end{equation}  
The effective field can be calculated using the following relation:
\begin{equation} \label{MM:002}
	\vec{H}_{eff} = - \left(     \dfrac{\partial E_{Tot}}{\mu_{o} \partial \vec{M}}           \right) \; \; \;\; \; \; \;\; \; \; \;\; \; \; \;\; \text{(in SI units)}.
\end{equation}
The micromagnetic form of various energy terms are described in the following.

\subsubsection{Self-Magnetostatic or Demagnetizing Energy}
 
Consider a FM thin film of finite size and  net magnetization $ \textit{M} $.  Every  magnetization component (at the film boundaries) normal to the surface gives rise to uncompensated fictitious magnetic charges (\textit{magnetic poles}), and the resultant field of these  charges is called the \textit{demagnetizing field}. The demagnetizing field ($\vec{H_{D}}$) acts in the opposite direction to the magnetization, $\vec{M}$. Thus, the magnetic induction inside the bar magnet after removing the external field becomes
\begin{equation} \label{MM:003}
	\vec{ B} = \mu_{o}(\vec{H} + \vec{M}) = -\mu_{o}\vec{H}_{D} + \mu_{o}\vec{M}  \; \; \;\; \; \; \;\; \; \; \;\; \; \; \;\; \text{(in SI units)}.
\end{equation}  
One can calculate the demagnetizing field (in micromagnetics) using the following expression:
\begin{equation} \label{MM:004}
	H_{Di} =  -N_{ij}M_{j},    \;  \; \; \; \; \;   \;  \; \; \; \; \;  \;  \; \; \; \; \;     i, j = x, y, z
\end{equation}  
where $N_{ij}$ is called the demagnetization tensor and  is usually written using a 3 x 3 matrix that has   unit  trace.  In the case of an ellipsoid, $\vec{H_{D}}$ and $\vec{M}$ are collinear along the principle axes, and the off-diagonal elements in the demagnetizing tensor are zero \cite{blundell2003magnetism,chikazumi2009physics,coey2010magnetism,cullity2011introduction,stancil2009spin}.  Thus,
$N_{x}+ N_{y}+N_{z} = 1$ (in SI units),  or   $N_{x}+ N_{y}+N_{z} = 4 \pi$ (in CGS units).
For an infinitely large thin film that lies in $xy$ plane, these demagnetizing factors become \cite{cullity2011introduction}
\begin{equation}\label{MM:004_1}
	N_{x} = N_{y} = 0, \;  N_{z} = 1.
\end{equation}
The energy associated with the  demagnetizing field is called the \textit{self-magnetostatic} or \textit{demagnetizing energy} and can be written as
\begin{equation} \label{MM:005}
	E_{Ms} = -\frac{\mu_{o}}{2} \int \vec{H}_{D}\cdot\vec{M}  = \frac{\mu_{o}}{2} \int  N_{D}M^{2}    \;  \; \; \; \; \;   \;  \; \; \; \; \;  \;  \; \; \; \; \;  (\text{SI units}).
\end{equation}     

\subsubsection{Exchange Energy}
The exchange energy is essentially short range, and it involves a summation over nearest neighbors. If we assume a small spatial variation of the magnetization at the mesoscopic length scale, the exchange energy density can be written as  \cite{stancil2009spin,vittoria2011magnetics,miltat2007numerical}
\begin{equation} \label{MM:006}
	E_{Ex} =\int\left( \dfrac{2A}{\mu_{o}M_{S}^{2}}   \nabla^{2}\vec{M}\right)dV,
\end{equation} 
where $ A   $ is the exchange stiffness constant in J/m (SI units). \\
\subsubsection{ Anisotropy Energy}
\textit{Anisotropy} refers to the directional dependence of magnetic material properties, playing a crucial role in the magnetic reversal mechanism and holding significant fundamental importance. 
 There are several  anisotropy energy terms relevant for thin film magnetism. A brief overview is given in the following.
 
\paragraph{Magnetocrystalline Anisotropy Energy:} The Magnetocrystalline anisotropy energy has its origin at the atomic level. In materials exhibiting significant anisotropy, there exists a robust coupling between the spin and orbital angular momenta within individual atoms due to spin-orbit coupling. Additionally, atomic orbitals, which are typically non-spherical, exhibit strong interactions with the lattice. Due to their shape, these orbitals tend to align preferentially along specific crystallographic directions. Consequently, the interplay between spin-orbit and orbit-lattice coupling dictates a preferred alignment for the magnetization, known as the easy direction. The energy required to deviate the magnetization from this preferential orientation is termed \textit{magnetocrystalline anisotropy energy}. Given the diverse lattice structures present in materials, it is natural to anticipate that this anisotropy energy is dependent on the lattice structure. 

\textit{Uniaxial Magnetocrystalline Anisotropy} occurs in hexagonal crystals, such as cobalt.  Phenomenologically one can write the energy expression as \cite{buschow2005concise}
\begin{equation} \label{MM:007}
 E_{Uniaxial} = K_{u}V sin^{2}\theta,
   \end{equation}    
where $\theta $ is the angle between the easy direction and the magnetization,  $K_{u}$ is the anisotropy constant in units of energy/volume, and $V$ is the volume of the sample.

\textit{Cubic magnetocrystalline anisotropy} occurs in cubic crystals, such as iron and nickel. 
Phenomenologically, one can write the energy expression as \cite{buschow2005concise,blundell2003magnetism,chikazumi2009physics,stancil2009spin}
\begin{equation} \label{MM:008}
 E_{Cubic} = K_{o}V + K_{1}V (       \alpha^{2}_{1}  \; \alpha^{2}_{2}  \;  +  \alpha^{2}_{2}  \; \alpha^{2}_{3}  \;  +     \alpha^{2}_{3}  \; \alpha^{2}_{1}  \;         )
   \end{equation}   
Here, the $\alpha_i$ represent the cosines of the angles between the magnetization direction and the crystal axes.

\paragraph{Shape Anisotropy Energy }
The shape anisotropy energy arises from the tendency of a FM to align its magnetic moments along certain preferred directions due to its shape. In the case of a two-dimensional nanowire, the demagnetizing factors are lower along the long axis compared to the short axis. This means that it is easier to magnetize the nanowire with an in-plane external field applied along the long axis, leading to an induced anisotropy known as shape anisotropy. The shape anisotropy energy increases as demagnetizing field  $\vec{H_{D}}$ increases.  A common phenomenological expression for the shape anisotropy energy for an ellipsoidal ferromagnet is given by:
\begin{equation} \label{MM:009}
E_{Uniaxial} = K_{D}V \sin^{2}\theta.
\end{equation} 
Here, $\theta $ is the angle between the long axis of the sample and the magnetization direction. The anisotropy constant is $ K_{D} =  \dfrac{1}{2}  (N_{x} - N_{z}) M^{2}$, where $ N_{x}$ and $N_{z}$ are demagnetization  factors in the short and long axis direction, respectively.

While various anisotropy energy terms exist, shape anisotropy dominates artificial spin ice dynamics. This preference arises because the patterned thin films from which ASIs are made are typically polycrystalline, with individual grains of the magnetic film (e.g., a NiFe thin film on a thermally oxidized Si substrate, with no perpendicular magnetic anisotropy and random grain orientation). In such a case, the influence of magnetocrystalline anisotropy is expected to be negligible compared to shape anisotropy.


\subsection{Micromagnetic Simulations }
Advances in computational hardware and numerical techniques enable the implementation of the micromagnetic approach described above on both high-performance computing  facilities and standard office desktop computers, utilizing either CPU or GPU processing.  

The key advantages of this micromagnetic simulations (MS) approach are as follows:
\begin{enumerate}
	\item \textbf{Replicating Experimental Observations}: MS accurately replicate experimental phenomena such as FM hysteresis and ferromagnetic resonance (FMR) spectra, providing a virtual laboratory for validation and exploration.  
	\item \textbf{Deciphering Parameter Influences}: By exploring the dependence of experimental outcomes (e.g., coercivity, resonance frequency) on simulation parameters like the exchange length of the FM, MS facilitates a nuanced understanding of these relationships, guiding experimental design and interpretation.
	\item \textbf{Optimizing Future Experiments}: MS empower researchers to fine-tune physical parameters for upcoming experiments, enhancing efficiency and efficacy by iteratively refining hypotheses and experimental setups.
\end{enumerate}
A finite difference method (FDM) based program called Object Oriented Micromagnetic Framework (OOMMF) can be used to perform MS \cite{donahue2002oommf}.  The geometry of an object needs to be first discretized in the form of cuboid spins of sizes \textit{p} nm $\times$ \textit{q} nm $\times$  thickness nm.  The magnitude  ($ \lvert \vec{M}\rvert $) of the magnetization vector within each cuboid is assumed to be constant.  The equilibrium magnetization [$ \vec{M} \equiv \vec{M}(\theta,\; \phi)$] state for a given geometry under the application of global external field is obtained using the gradient descent method (GDM).  The GDM method is based on the assumption that a physical system will evolve along a path that minimizes the total energy of a given system. 

The above stipulated process is iterated until the $\hat{M }\times (\hat{M }\times \vec{H}_{eff}) = \vec{B}_{eff}/\mu_{o} \;< \;0.1 \;A/m$,
where the total internal effective field can be obtained by using the formula 
\begin{equation}
	\vec{H}_{eff} = - \left(     \dfrac{\partial E_{Tot}}{ \partial \vec{M}}           \right) \; \; \;\; \; \; \;\; \; \; \;\; \; \; \;\; \text{(in SI units).}
\end{equation} 
OOMMF provides functionality for storing both local and global magnetization values, as well as local demagnetization field values. These local quantities can be visualized as a function of spatial coordinates using the OOMMF graphics viewer module or the open-source OOMMF Tools module \cite{OOMMFtools1,donahue2002oommf}.   
Dynamic magnetization simulations can be performed using OOMMF in conjunction with user-developed postprocessing scripts. The steps for calculating the FMR frequency under a given external magnetic field are as follows:
\begin{enumerate}
	\item Determine the equilibrium magnetization configuration for a given external field using the energy minimization method described earlier.
	\item  Introduce a Gaussian magnetic field pulse perpendicular to the film plane, with a full width at half maximum (FWHM) of a few picoseconds (e.g., 2.5 ps) and an amplitude of a few millitesla (e.g., 5 mT). After applying the pulse, the system relaxes dynamically according to the LLG equation. A map of the local magnetization can be then recorded in $\Delta t$ steps (for example 20 ps) for a total time duration of \textit{T}.   
	The frequency resolution of the simulated ferromagnetic resonance spectra can be written as $\Delta f = \dfrac{1}{T}$. The stored data for    the magnetization $\vec{M}\equiv \vec{M}(x,y,z,t_{i})$ can then be read out into  \textit{MATLAB} matrix or Numpy format  \cite{OOMMFtools1} and Fourier transformed. The local power density, (i.e., the squared   amplitude for each pixel)  can be integrated over the  whole geometry to yield the global power density.   Also, the local power densities can be mapped for each    frequency to show the inhomogeneous distribution of   the power absorbed  for the corresponding frequency.    
\end{enumerate}
Apart from OOMMF, other open-source (e.g., Mumax3 \cite{vansteenkiste2014design}) and proprietary micromagnetic simulations packages \cite{leliaert2019tomorrow} have been successfully utilized for simulating FM nanostructures and interested readers are referred to Ref.~[\onlinecite{leliaert2019tomorrow}] for a list of these simulation packages.


\section{Nanofabrication}\label{Nanofab}
Fabricating sub-micron 2D and 2.5D ASIs for magnetization dynamics measurements, composed of metallic magnetic films like Ni\textsubscript{81}Fe\textsubscript{19}, often involves positioning them at a nominal distance of 50 nm or less from each other. Techniques such as electron beam lithography (EBL) \cite{jungfleisch2016dynamic,bhat2016magnetization} and deep ultraviolet lithography \cite{zhou2016large} with lift-off methods are commonly employed for the fabrication of these intricate nanostructures. Of these, EBL with lift-off stands out as a preferred method, primarily for its ability to create prototypes of arbitrary periodic and aperiodic shapes without the need for mask blanks, thus expediting the fabrication process. Moreover, EBL facilities are widely accessible at research universities, further enhancing its appeal.
\par
\begin{figure*}
\includegraphics[width=\textwidth]{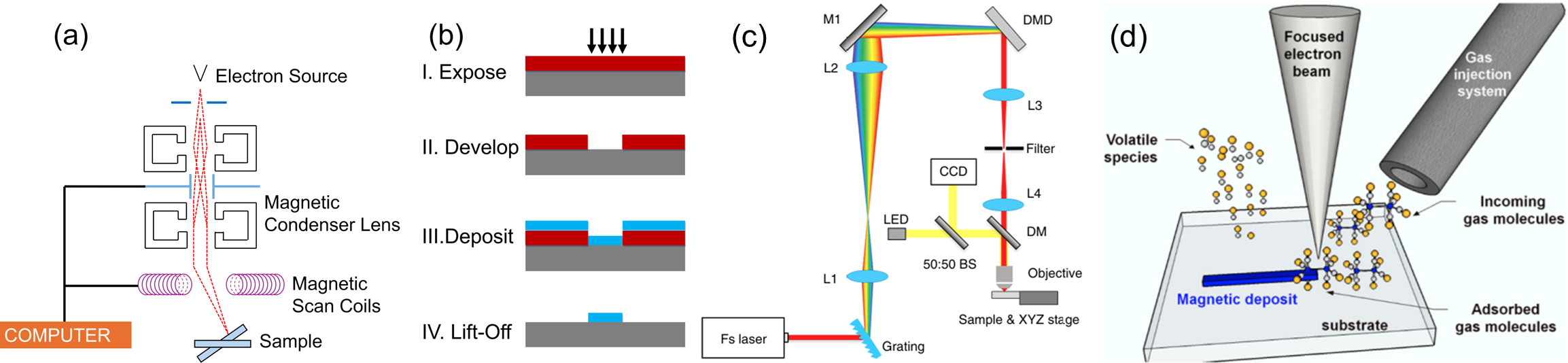}
	\centering
	\caption{(a) Conversion of scanning electron microscope (SEM) into an electron beam lithography (EBL) tool showcasing control of deflection coils and blanking elements. (b) Nanofabrication procedure using positive resist and lift-off technique. (c) The optical configuration of the two-photon (TPL) lithography fabrication system utilizes femtosecond laser beams, which are scanned using the digital micro-mirror device (DMD) multi-point random-access scanner. \cite{geng2019ultrafast} (d) A schematic representation of focused electron beam induced deposition (FEBID) is shown. FEBID is an electron-assisted chemical deposition technique that enables high-resolution patterning in a single step. \cite{de2016review}}\label{Figure_3}
\end{figure*}

EBL systems utilized in the fabrication of ASI samples are typically either converted scanning electron microscope (SEM) systems [Fig. \ref{Figure_3}(a)] or standalone EBL systems. Converted SEM systems are typically limited to a 30 keV acceleration voltage, whereas standalone EBL systems routinely operate at 100 keV. Standalone EBL systems capitalize on higher accelerator voltages to achieve finer resolution  in thicker resists by exploiting reduced forward scattering and longer-range back scattering for higher primary electron energies\cite{manfrinato2014determining}.  
Notably, bilayer resist-based lift-off processes often exhibit lower resolution compared to single-layer resist techniques \cite{madou2018fundamentals,jaeger1987introduction}.
\par
The initial phase in crafting magnetic nanostructures involves applying a mask that responds to electrons of specific energies from the EBL instrument. This mask, known as the e-beam resist (ER), comprises a solid polymer that is typically dissolved in a liquid solvent for spin coating and is sensitive to the incoming electron beam. Upon interaction with the ER, the primary electron, or secondary electrons, either excites or ionizes atoms in the polymer.  These processes trigger chemical reactions within the ER, leading to its categorization as either positive or negative ER, based on the beam's effect. For positive ER, the incident beam prompts chain-scission reactions, breaking polymer chains into smaller fragments, thus, reducing the molecular weight and enhancing solubility in the developer solvent. Conversely, negative ER undergoes cross-linking reactions, increasing molecular weight and reducing solubility in the developer. Consequently, positive ER dissolves in exposed areas, leaving behind the unexposed regions to form the desired pattern \cite{chen2015nanofabrication,gangnaik2017new}.\par

The performance of an ER is evaluated based on several parameters:
a)	Tone: Dictates the portion of ER removed during development, determining whether the ER is positive or negative.
b)	Sensitivity: Determines the required dose required to expose the ER.  It is typically measured in primary electron charge per area but can also be measured in energy per volume.  
c)	Resolution: Specifies either the minimum feature size or the minimum pitch achievable with a particular resist.
d)	Contrast: Describes the change in the rate of resist development with dose.
e)	Etch Resistance: Determines the ER's resistance to chemical or physical etching.
Two widely used positive ERs for the fabrication of ASI include:

\textit{Polymethylmethacrylate (PMMA):} Renowned for its high resolution (2-20 nm)\cite{manfrinato2017aberration}, PMMA exhibits high contrast but low sensitivity and resistance to dry etching \cite{chen2015nanofabrication}. Due to its low sensitivity, PMMA demands a higher dose for effective clearing.  A bilayer resist approach is often used with PMMA in which a combination of high and low molecular weight PMMA generates an undercut during development, aiding the lift-off process.

\textit{ZEP:} ZEP ER \cite{gangnaik2017new} boasts three times the sensitivity of PMMA, offering similar resolution while reducing EBL writing time. Additionally, a single layer of ZEP achieves the desired undercut when exposed at 1.5 times the clearing dose, streamlining the lift-off procedure and obviating the need for a bilayer resist. Another viable option is CSAR-62, a chemically semi-amplified resist \cite{schirmer2013chemical}, thrice as sensitive as PMMA and more cost-effective than ZEP, making it a feasible choice for ASI fabrication.

The key process steps routinely employed to obtain magnetic nanostructures are outlined below [Fig. \ref{Figure_3}(b)]:
\begin{enumerate}
\item Spin coat an ER onto a substrate (e.g., silicon wafer).
\item	Expose the ER using EBL.
\item	Develop the ER in an appropriate solution.
\item	Deposit a magnetic thin film (typically 3-25 nm thick) or multilayer onto the patterned substrate.
\item	Strip off the remaining resist by immersion in an appropriate chemical solution.
\end{enumerate}

\subsection{Other approaches for fabricating ASI, including 3D ASI:}


\textbf{Two-Photon Lithography (TPL):}

Recently, two-photon lithography (TPL) has become one of the most widely used high-resolution techniques for fabricating intricate 3D structures with nanoscale precision, making it an ideal method for creating 3D ASI samples \cite{nano10030429, Guo2023Rea, Williams2018Two, may2021magnetic, pip2022x}. The methodology of TPL involves the following key steps [Fig. \ref{Figure_3}(c)]:

\textit{Preparation of the photopolymer resist:}
A photosensitive resist is chosen that allows fine-resolution polymerization under two-photon absorption. The resist typically contains monomers and photoinitiators that react to femtosecond laser pulses. The material needs to be compatible with the deposition of magnetic material in subsequent steps.

\textit{Design of the spin ice structure:}
The ASI design involves a lattice of magnetic islands that act as individual ``spins''. These islands are arranged in a predefined 3D grid pattern. CAD software is used to design the geometry of the islands and their spatial arrangement, ensuring the structure mimics the characteristics of a spin ice material.

\textit{Laser system setup:}
A femtosecond laser with an appropriate wavelength (800 nm – 1030 nm, typically Ti:Sapphire) is selected to induce two-photon absorption in the resist. The laser is focused through a high-numerical aperture objective lens to create a highly localized polymerization spot. By scanning the focused laser across the resist, the desired 3D pattern of the spin ice structure is written, one layer at a time.

\textit{Two-photon polymerization:}
The laser pulses induce two-photon absorption only at the focal point, where the light intensity is sufficiently high, leading to localized polymerization. Owing to the nonlinear nature of this process, polymerization is confined exclusively to that focal region. By precisely controlling the movement of either the laser beam or the sample stage in three dimensions (X, Y, and Z), the structure is written point by point or line by line within the resist, allowing for the fabrication of high-resolution polymerized 3D islands that form the ASI lattice. The laser intensity, scanning speed, and exposure duration are controlled to define the dimensions of each island and ensure accurate positioning.

\textit{Post-exposure development:}
After the microstructure is written, the sample undergoes a development process where the unpolymerized resist is washed away, leaving behind the polymerized 3D structure. The sample is then dried using critical point drying (CPD) to prevent structural collapse.

\textit{Magnetic material deposition:}
To create the artificial spins, the polymerized 3D structure is coated with a thin layer of ferromagnetic material such as Co or NiFe using techniques like sputtering or electrodeposition. Nickel can also be deposited by atomic layer deposition to provide excellent conformal coverage\cite{guo2023realization}. A lift-off process (if needed) is used to remove excess metal, leaving behind the ASI nanostructures. The magnetic islands are then magnetized to establish their magnetic dipoles.

\textit{Characterization:}
The final 3D ASI sample is analyzed using SEM for structural verification and magnetometry to assess the magnetic properties, ensuring the islands behave as individual magnetic ``spins''.

\textit{Post-fabrication processing (optional):}
Depending on the application, post-fabrication processing, such as annealing or additional chemical treatments, may be performed to improve the mechanical properties or to further refine the structure. This can also include surface functionalization for specific applications in areas such as biosensors or photonic devices.

\textbf{\textit{Advantages of Two-Photon Lithography:}}

\textit{High Resolution:} TPL achieves sub-micron resolution, with the ability to fabricate features as small as 100 nm, significantly below the resolution of traditional single-photon lithography at comparable wavelengths.

\textit{Three-dimensional structuring:} Unlike conventional photolithography, which is limited to two-dimensional patterns, TPL can create complex 3D structures in a single step. In two-photon polymerization (2PP), regions outside the laser focus are less likely to reach the polymerization threshold of the photoresist. This characteristic enables the creation of complex 3D structures, as the proximity effect in two-photon absorption is much weaker compared to that in one-photon absorption.

\textit{Localized polymerization:} The two-photon absorption process ensures that polymerization occurs only at the focal point of the laser, reducing the risk of unwanted material solidification outside the desired region.

$\newline$

\textbf{Focused electron beam induced deposition (FEBID):}

Another technique that has recently emerged as a promising way to create ASI is focused electron beam induced deposition (FEBID), which is a vacuum-based nanofabrication technique that writes 3D nanostructures using a high-energy focused electron beam to dissociate molecules on a substrate, regardless of their geometry, without the need for resist or solvents \cite{Thorman2015, Berchialla2024Focus, meng2021non, skoric2022domain}. This technique involves the use of a focused electron beam to induce chemical reactions that locally decompose a precursor material, allowing for the controlled deposition of material to form complex 3D structures [Fig. \ref{Figure_3}(d)].

\textit{Preparation of substrate and precursor material:}
The substrate, often a silicon wafer or conductive material, is cleaned and prepared to support the deposition process. The electron-sensitive precursor, typically an organometallic compound is introduced as a gas an adsorbs to the substrate.  The precursor will undergo decomposition under the influence of the electron beam, enabling the deposition of metallic magnetic islands.  For magnetic materials the precursor is typically dicobalt octacarbonyl, iron pentacarbonyl, or one of a variety of nickel precursors.\cite{magen2021focused, escalante2023long}  Deposition from condensed liquid precursors is also possible including nickel.\cite{donev2016nanoscale} 

\textit{Design of ASI lattice:}
The 3D ASI structure is designed using specialized CAD software. The design includes a carefully arranged lattice of magnetic islands, which will behave like individual spins in the ASI system. This design dictates the spatial arrangement, size, and orientation of each magnetic island.

\textit{FEBID process setup:}
The FEBID process takes place within an SEM system equipped with a focused electron beam. The electron beam is finely focused to direct energy onto the precursor material, causing localized deposition at specific locations.

\textit{Electron beam scanning and deposition:}
The electron beam is scanned over the substrate according to the designed lattice. The energy from the beam decomposes the precursor material, selectively depositing metallic material at the targeted positions to form magnetic islands. The intensity, dwell time, and scanning speed are precisely controlled to achieve accurate deposition and structural integrity.  This often requires precise modeling and correction of the growth process to achieve true 3D structures.

Other possibilities for nano-structuring include 3D printing and self-assembled templates to fabricate 3D nanostructures. More details about these recent development of 3D ASI fabrication can be found in Refs.~[\onlinecite{nano10030429, Berchialla2024Focus, 3D_roadmap}].

\section{Microstate control}
\label{microstate}

In the previous sections, we discussed the theoretical background and common fabrication methods of artificial spin ice systems, including comments on the vast microstate spaces characteristic of ASI. The specific microstate defines the dynamic response of the ASI, hence the ability to prepare specific microstates is of great interest. In the following section, we discuss approaches for controlling ASI microstates at a global and local level.



The majority of ASI research focuses at least in part on the system microstate -- whether with fundamental physical motivations such as statistical studies of vertex populations and avalanche dynamics, or functional studies where the microstate defines system performance, such as in reconfigurable magnonics or neuromorphic computing. As such, methodologies for ASI microstate control are of high importance.

Here, we will cover established microstate control approaches and comment on horizons for future development. Microstate control and the vast reconfigurable microstate space of ASI is at the very core of what makes magnetization dynamics interesting in these strongly-coupled nanomagnetic array systems, so we have described the fundamental concepts and applied techniques in detail.

\subsection{Global vs Local Microstate Control Techniques}

Broadly speaking, control of the ASI microstate can be achieved either by global techniques which address all islands simultaneously by applying some magnetic field or thermal protocol across the whole system, or local techniques in which the magnetization of individual islands is directly addressed. 

Global techniques have the advantage of simplicity (the most common global control technique is the application of a magnetic field) but generally can only achieve control of the ASI microstate at a statistical average level, where the net ratio of different vertex populations is controlled but not the spatial distribution of the different vertex types throughout the system. Because of the vastly degenerate energy landscape of ASI systems it is generally not possible to access the full microstate space using such global techniques, which tend to favor lower energy vertex configurations.

In contrast, local techniques which use either scanning probes, optical pulses, local microwave or magnetic fields from striplines/waveguides or electrical control to switch islands may be fully deterministic and can in principle be used to access any microstate of an ASI array. The cost of this greater control is that `writing’ the microstate of an ASI system island-by-island typically takes significantly longer than addressing all islands simultaneously, and requires substantially harder device engineering. An ASI system limited to a single $100 \rm{\mu m} \times 100 \rm{\mu m}$ EBL writefield can easily contain $10^5$ to $10^6$ islands, so that to keep the total array writing time below one minute -- each island would have to take no longer than $10 ~\rm{\mu s}$ to address. This is out of reach for most scanning probe systems and at the limit of what might be achieved using optical methods with fast-steering mirrors, though recent advances in spatially structured light are appealing here\cite{forbes2021structured}.

There is some scope for improving write speeds by addressing multiple islands simultaneously or wiring up electrical address lines to every island in the array, but in the short term approaches tend to be local, precise, slow and experimentally challenging -- or global, coarse, typically somewhat stochastic/statistical and experimentally easy. Combined local/global approaches are appealing, where a small number of nanoislands may be locally written before applying global fields to `seed' some initial conditions of microstate evolution.

\begin{figure*}
\includegraphics[width=0.9\textwidth]{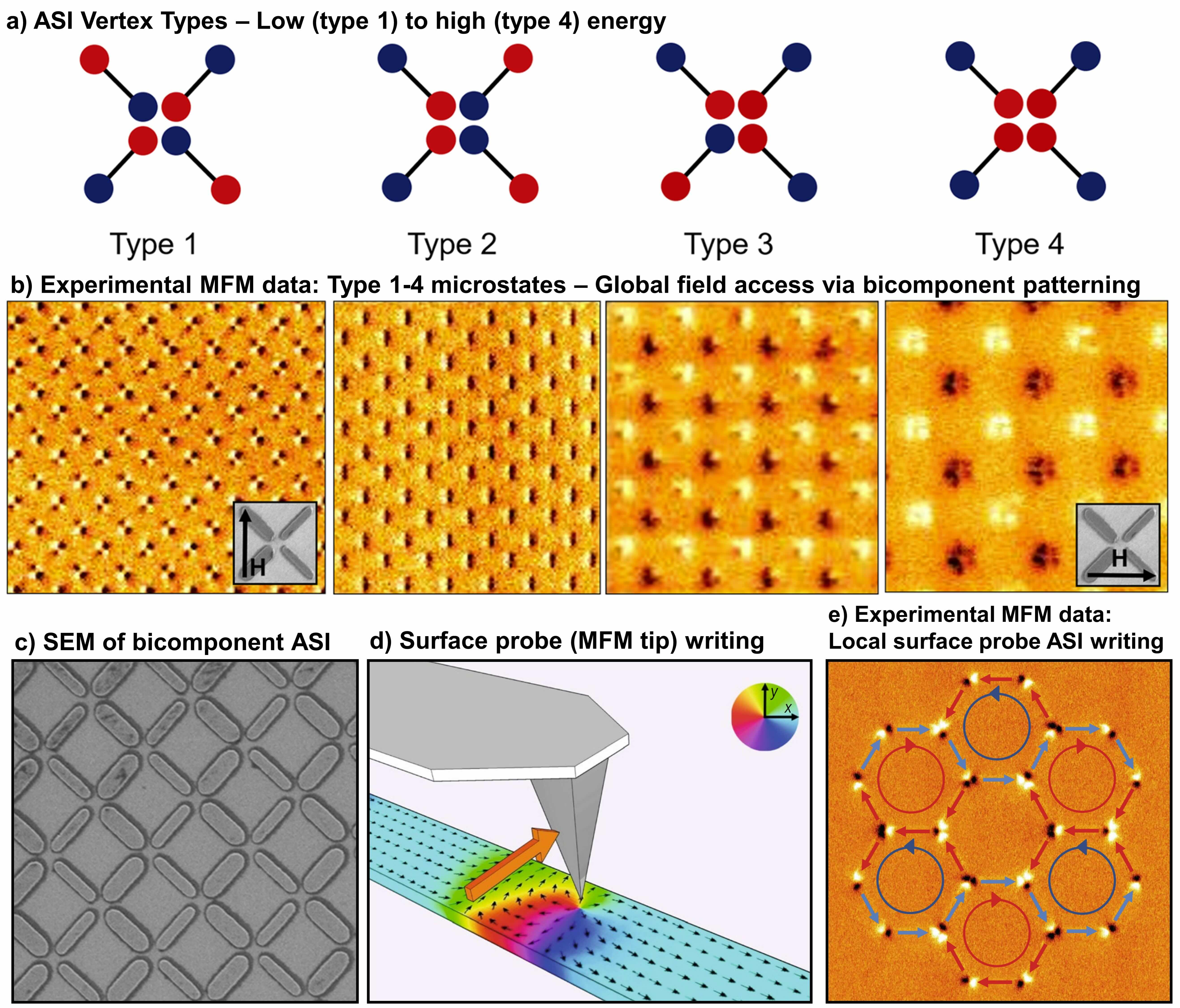}
	\centering
	\caption{ASI microstate control, global field and local scanning probe approaches. (a) The four possible vertex types in square artificial spin ice illustrated schematically as macrospins/charge dumbells. Red and blue circles represent positive and negative magnetic charge (e.g., north and south poles), respectively. System energy increases from type 1 (low energy) to type 4 (high energy). (b) Experimental magnetic force microscope data showing ASI arrays prepared in type 1-4 vertex states, corresponding to the schematics in (a). Here, global field control is used combined with bicomponent nanopatterning where an ASI array of wide (low coercive field) and thin (high coercive field) has been fabricated to allow for expanded microstate control using simple global field control. Adapted from Ref.~[\onlinecite{gartside2021reconfigurable}]. (c) Scanning electron micrograph illustrating the bicomponent nanopatterning allowing the expanded global field microstate control shown in (b). Adapted from Ref.~[\onlinecite{gartside2021reconfigurable}]. (d) Schematic of local microstate writing via surface probe, using a magnetic force microscope tip. The local magnetic field from the tip apex injects vortex core topological defects to nanoislands, resulting in magnetization reversal of the macrospin state of the nanoisland. Adapted from Ref.~[\onlinecite{gartside2018realization}]. (e) Experimental magnetic force microscope data demonstrating a complex ASI microstate written via local surface probe control. Adapted from Ref.~[\onlinecite{gartside2018realization}].}\label{Microstate_figA}
\end{figure*}

\begin{figure*}
\includegraphics[width=0.9\textwidth]{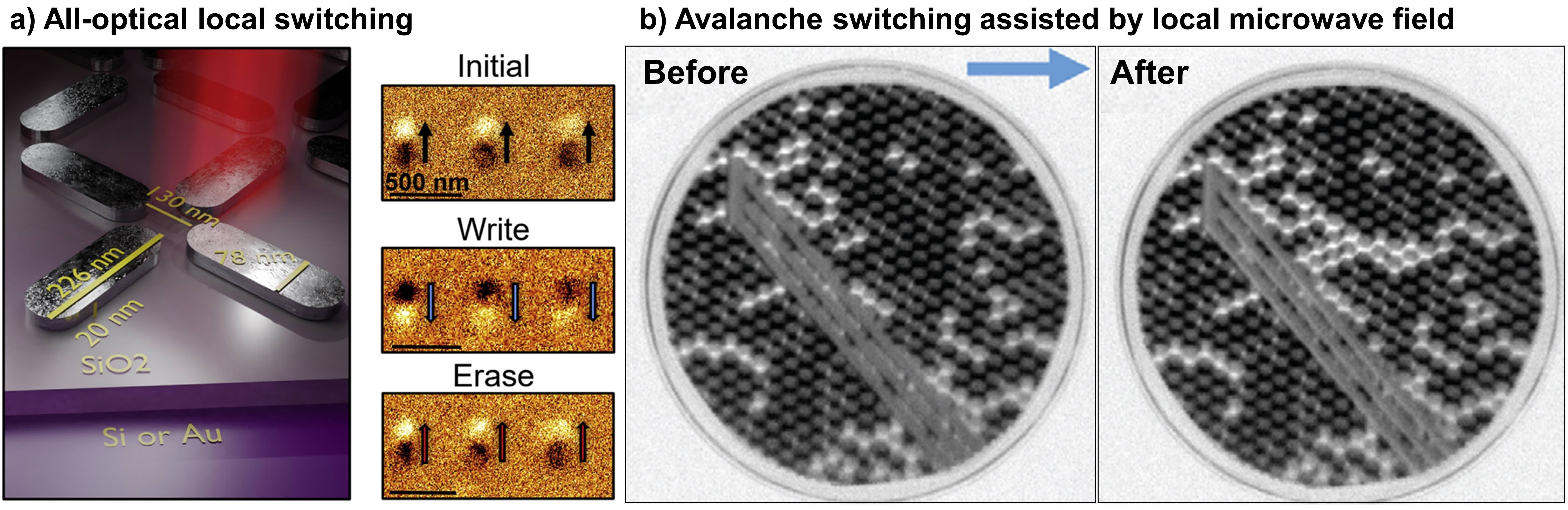}
	\centering
	\caption{ASI microstate control, local optical writing and microwave assisted control. (a) All-optical writing of ASI, allowing local control. Write/erase toggle switching is possible. Adapted from Ref.~[\onlinecite{stenning2023low}]. (b) Avalanche switching in ASI assisted by microwave field supplied by micro-patterned waveguide. Adapted from Ref.~[\onlinecite{bhat2020magnon}].}\label{Microstate_figB}
\end{figure*}

\subsection{Global Control}\label{static}
\subparagraph{Magnetic field protocols (effective temperature):}$ \newline  $
A simple method to generate a reproducible state of all `Type II' vertices [e.g., field polarized, see Fig.~\ref{Microstate_figA} (a) and (b)] in a square-lattice ASI system might involve applying a magnetic field above the nanoisland coercive fields so that the whole array is saturated along the field axis. This approach is quick and reliably `deletes’ the memory an ASI sample has of its microstate history, but how can magnetic fields be used to access a broader range of microstates than just the saturated state?

For each island in an ASI array there is some coercive switching field $H_{i}^{\rm switch}$ at which that island’s magnetization will reverse. If the islands in a system are identical, then they will all reverse at the same external applied field strength except where local dipolar magnetic fields from the nanoislands locally effectively reduce or increase the global field magnitude required to switch nanoislands. Such local modulations in the magnetic field may occur due to stray fields from neighboring islands as in the case of artificial ferromganetic quasicrystal tiling \cite{bhat2013controlled}, but in the example above all vertices have the same configuration after the saturation field has been applied and hence all islands are subjected to identical fields from neighbouring elements. If the system starts in a random configuration before the saturation field is applied then it may pass through some intermediate microstates as the external field is ramped up to the coercive field, but once a system of identical islands enters a state where all vertices have the same spin configuration any further field looping will generally only be able to access a small set of microstates.

Thankfully, imperfections in fabrication mean that the islands in experimental ASI systems are generally not identical but rather have some distribution of switching fields. Empirical work on a range of ASI systems \cite{Shen2012, Huegli2012, Morgan2013} indicates that the effect of these imperfections, referred to as `quenched disorder’, can generally be modeled as a Gaussian distribution of switching fields $H_{i}^{\rm switch} \sim \mathcal{N}(\mu,\sigma^2)$ where $\mu$ and $\sigma^2$ are respectively the mean and variance of the distribution. The moments of the distribution can be estimated by measuring a hysteresis loop and fitting a Gaussian to $\partial M / \partial H$ as a function of the applied field $H$. Though strictly speaking such a measurement does not take into account the effect of local stray fields, in practice the effect of quenched disorder typically dominates the effect of dipolar inter-island interactions for nanoscale magnetic arrays \cite{Fraleigh2017, Kempinger2020, Vanstone2022}.

The randomization injected by quenched disorder into the microstate trajectory of ASI systems responding to some magnetic field sequence massively expands the set of microstates accessible by applying global fields, allowing islands with low switching fields to act as nucleation points located at random sites in the lattice \cite{Budrikis2011}. For fields sequences in which the field strength is kept below the level that forces all macrospins to align with it (known as `minor loops’) the strong path dependence of microstate trajectories leads to a particularly rich ensemble of branching dynamical pathways~\cite{jungfleisch2016dynamic} which can be used to provide the `fading memory’ required for reservoir computing \cite{Gilbert2015, Carroll2022,gartside2022reconfigurable}. Since for most early ASI systems the energy barrier of $K_{\rm D}V$ associated with shape anisotropy [from Eq. (\ref{MM:009})] was orders of magnitude greater than the thermal energy scale $k_{\rm B} T$, such ASI systems were `athermal' in the sense that thermal fluctuations were generally incapable of reversing the magnetization direction of individual islands, preventing the spin state from reaching its equilibrium configuration. As a result, a large body of work has explored whether quenched disorder can take the place of random switching due to thermal fluctuations to direct ASI systems into ground state configurations whose statistical properties match those of systems in thermal equilibrium. Such configurations are of interest where the focus is on thermodynamic properties of ASI systems or on the behaviour of low-energy excitations such as monopoles.

The general strategy for achieving low-energy states is to use alternating or rotating magnetic fields whose magnitude is gradually decreased so that island reversals that lower the magnetic energy are favored \cite{wang2007}, analogous to a thermal anneal. An analogy with thermal processes has proved useful in describing the statistical properties of the magnetic configurations achieved through these field protocols, for which the number of different vertex types after certain rotating field protocols were found to follow a Boltzmann distribution with an effective temperature \cite{Nisoli2010} whose value varies linearly with the size of the steps with which the field strength is reduced and decreases in proportion to the strength of the inter-island interactions. This effective thermodynamics picture has been extended to describe the low-energy states that can be accessed by field sequences \cite{Nisoli2007, Lammert2010, Lammert2012} but while such sequences have been shown to reliably generate low-energy magnetic configurations with small net moments, it seems that even in the limit of very small field steps rotating field protocols cannot access the ground state \cite{Ke2008, Morgan2013}.

This failure to reach true thermal states can be explained using a network model \cite{Budrikis2012a, Budrikis2014} where microstates are represented as nodes connected by edges where a transition between two microstates is possible through a single spin flip \cite{farmer2016direct} for which the corresponding switching field is below the local magnetic field strength. It can be shown within this model that although quenched disorder vastly increases the number of microstates that can be reliably accessed, non-random field sequences (e.g., linear field sweeps or rotating field protocols) generally tend to end up in dynamical limit cycles localized in a strongly connected component of the wider microstate network. This is a result of the fact that athermal ASI systems can only ever lower their energy in response to applied magnetic fields, unlike thermally active ASI systems in which thermal fluctuations may increase the internal energy to overcome energy barriers, which prevent access to the ground state. 

Both the effective temperature model and the network model are useful to explore general results about how the ASI microstate can be controlled with applied magnetic fields, but for specific systems it can be useful to directly simulate the response to an external field. 
Carrying out full micromagnetic simulations is generally too costly for treating large arrays.  Alternatives such as network models simply track whether local fields exceed the switching fields of each island and reverse their magnetizations accordingly.  This approach can be used to efficiently track which microstates can be accessed by field protocols. The flatspin \cite{Jensen2022} package efficiently carries out simulations of ASI systems using this approach.


\subparagraph{Thermal protocols (quenching, annealing, phase transitions, as-grown):}$ \newline  $
The other major statistical approach to preparing reproducible ASI configurations is through thermal techniques. While the rotating field protocols discussed above were motivated by the fact that early ASI systems were athermal, it has since become possible to fabricate thermally active systems.

The first hint that it might be possible to create thermal ASI systems came from the observation that vertex configurations in `as-grown’ samples obey Boltzmann statistics \cite{Morgan2011, Greaves2012, Nisoli2012}. This was interpreted as a result of the system being thermally active during the deposition process while the islands had not yet reached their full thickness (and hence following Eq.~(\ref{MM:009}) the energy barrier $K_{\rm D}V$ was lower) and the thermal state being frozen in as the thickness increased. Since this initial observation, various means of activating thermal fluctuations in ASI have been discovered.

To access ground states, ASI systems can be thermally annealed by heating them to either the Curie temperature of the material used to form the islands or to the blocking temperature of the islands $T_{\rm b} \sim KV_{\rm D}/k_{\rm B}$ (it is worth noting that while the thermal energy barrier is often taken to be exactly $K_{\rm D}V$, careful modeling of the transition between the two microstates during switching suggests that it can be significantly lower than this \cite{Liashko2017, Koraltan2020, Leo2021}).

The Curie temperature for NiFe is around 850 K, which is significantly lower than the blocking temperature for ASI systems with thicknesses greater than 5 nm. However, heating permalloy ASI structures to this temperature has been observed to cause a substantial reduction in saturation magnetization\cite{zhang2013}. This effect has been attributed to the degradation of the permalloy islands, due to the interdiffusion of the permalloy into the Si/SiO$_{2}$ substrates typically used \cite{julies1999,schulz2021}. 

It is possible to circumvent this by lowering the Curie temperature and this was demonstrated with systems made of monolayers of Fe delta-doped into palladium islands \cite{Kapaklis2012, Arnalds2012} and with alternative materials such as Ni-rich permalloy \cite{Porro2013} and FePd alloys \cite{Drisko2015,Morley2018}. On the other hand, annealing ASI systems based on standard permalloy is possible if a silicon nitride substrate is used \cite{zhang2013,zhang2019} to prevent the permalloy diffusing into the substrate.

Alternatively, the blocking temperature can be lowered to below room temperature by reducing the thickness of the islands to below $\sim 5$ nm (the exact thickness threshold is system-dependent, so a common trick to maximize the chance that some region of a sample is thermally active is to deposit permalloy in a wedge so that the thickness varies over the system). This offers full thermal control over the vertex fractions, as well as the possibility of controlling the magnetic configuration by sweeping the temperature through phase transitions \cite{Moller2009,Levis2013,Anghinolfi2015, Sendetskyi2019}. By tuning the energy scales such that the critical temperature for a phase transition is close to room temperature,\cite{chen2019spontaneous} it could thus be possible to engineer a dramatic change in the spin configuration with small temperature modulations.

\subparagraph{Lattice modification:}$ \newline  $
So far, we have considered homogeneous arrays of regular ferromagnetic nano-islands, but a powerful means of expanding the range of microstates which can be accessed both through field protocols and thermal relaxation is to modify the lattice in a way that locally modifies interaction energies. While this generally breaks the degeneracy of ASI systems, the reward is ready access to specific subsets of microstates.

One strategy takes the idea of quenched disorder, where different shapes and sizes of nanoislands gives different switching fields, and directly modifies the switching field distribution by making some islands larger or smaller\cite{dion2019,gartside2021reconfigurable} or composed of different magnetic materials\cite{lendinez2021emergent} (e.g., CoFe and NiFe) such that they reliably reverse at different magnetic field amplitudes. This allows access to an enhanced range of microstates via simple global field. An example is a bicomponent square lattice ASI with thin and wide bars arranged in a `staircase' orientation that can lead to all four ASI vertex states\cite{gartside2021reconfigurable}.
This approach is illustrated in Figs.~\ref{Microstate_figA}(b-c).  First, the array is globally saturated.  Then, global magnetic fields are applied, such that only the wide bars reverse, which allows access to the four type 1-4 vertex states.  The type 1 ground state is attained by applying the second wide-bar reversal field along the vertical columns of wide bars.  The high-energy type 4 `monopole' state is attained by applying the second wide-bar reversal field horizontally (in the $x$-direction), perpendicular to the columns of wide bars. A slightly different approach modifies the switching field distribution, not by altering the nano-islands themselves but by applying localized fields to offset the global field; this has been achieved by the introduction of localized exchange bias fields \cite{parakkat2021}.


A final approach leaves the switching field of the islands unperturbed, but attempts to steer the magnetization dynamics by controlling the topology or boundary conditions of the ASI system. This might take the form of `seeding' monopole nucleation sites through the introduction of defects \cite{Drisko_2017} or nucleation sites at the system boundaries \cite{arava2020}. In connected ASI structures, the direction in which monopoles can propagate through the ASI systems once they have nucleated may further be constrained by controlling the vertex connectivity and topology \cite{Pushp2013,Singh2024,Arava2024}, which may be tuned so that domain walls propagating through vertices follow either deterministic or stochastic trajectories.

While many lattice modifications have been demonstrated to be effective as proofs-of-concept, this approach has not been investigated as thoroughly as the techniques based on magnetic and thermal field protocols. 
Lattice modification offers direct control over which microstates can be accessed while retaining the ability to rapidly switch between different configurations by simultaneously addressing all islands with a global field.  Therefore, exploring lattice modification as a means of systematic microstate control would be a promising avenue to a simplified and fast  ASI control.


\subsection{Deterministic Control}
\label{deterministic}
While the global field compatibility of the methods specified so far is attractive, the catch is that the accessible microstates are still hard-coded at fabrication, limiting flexibility. We will now turn to deterministic methods, which, in principle allow any microstate to be accessed.

To control the microstate of an ASI system deterministically, a means of locally reversing the magnetization of individual islands is required. This has been demonstrated in three ways so-far: surface-probe writing\cite{wang2016rewritable,gartside2018realization,lehmann2022poling}, optical control\cite{pancaldi2019selective,gypens2022thermoplasmonic,stenning2023low} and electrical control of anisotropy via magneto-ionics\cite{yun2023electrically}.

Surface-probe control has so-far used the stray dipolar magnetic field of a magnetic force microscope tip (see Sec.~\ref{Static}) to locally manipulate the ASI microstate. The first demonstration of this approach combined the local tip magnetic field with a globally applied field, such that away from the tip the global field was below the nanoisland coercive field, but close to the tip the sum of the global and the tip fields was sufficient to drive magnetic reversal\cite{wang2016rewritable}. After this, it was shown that a sufficiently strong magnetic tip could induce topological defects in the magnetic structure of the nanoisland, driving magnetic reversal as the tip scans over the nanoisland with no requirement for global magnetic field\cite{gartside2018realization}. This field-free approach is illustrated schematically in Fig.~\ref{Microstate_figA}(d), with an example of the complex ASI microstates accessible with such local control shown via experimental MFM data in Fig.~\ref{Microstate_figA}(e). The latter method was used to locally manipulate ASI arrays into a magneto-toroidal state\cite{lehmann2022poling}. To overcome the limited speed at which a magnetic force microscope tip can be moved between islands, it has been proposed that the localized magnetic field used to trigger magnetic reversal could be provided by the stray field of a domain wall driven through a ferromagnetic wire by a spin-polarized current \cite{Gartside2020}. Another possibility might be spin torque switching of particular islands or subsets or islands. Nevertheless, to date magnetic writing techniques based on scanning localized magnetic fields over islands remain too slow for application to large numbers of islands.

Similar to surface probe writing, optical control of ASI has been demonstrated with and without the need for additional global magnetic fields. The first example used a focused laser beam to plasmonically heat selected nanoislands, reducing their coercive field such that they could be switched by a global magnetic field while leaving other nanoislands unaffected\cite{pancaldi2019selective,gypens2022thermoplasmonic}. Another approach was demonstrated without the need for global magnetic fields, using just a focused laser beam to all-optically toggle-switch the magnetization state of NiFe ASI nanoislands\cite{stenning2023low} -- illustrated in Fig.~ \ref{Microstate_figB}(a). This was a somewhat surprising result as all-optical magnetic switching techniques typically need more complex magnetic materials such as ferrimagnetic alloys\cite{le2015nanoscale} or multilayered materials\cite{igarashi2023optically}, due to switching mechanisms which rely on different demagnetization speeds in different sublattices within alloys\cite{le2015nanoscale} or optically-induced spin transfer between layers\cite{igarashi2023optically}. Purely thermal optical switching (e.g., optically-induced heating and subsequent demagnetization) was shown to be an unlikely explanation in the NiFe ASI study\cite{stenning2023low}  due to the repeatable toggle switching and non-random fidelity of the demonstrated written states, including long chains of specific vertex types.  Further investigation is needed to reveal the physical mechanism responsible for the all-optical switching in NiFe ASI.

Microwave-assisted switching in ASI has been demonstrated. But is not yet fully deterministic.  Local control
was demonstrated using a weak global magnetic field combined with local RF excitation to reverse nanoislands
close to the RF antenna\cite{bhat2020magnon} [illustrated in Fig.~\ref{Microstate_figB}(b)]. This exciting direction has clear compatibility with the development of reconfigurable magnonics in ASI.


\section{Experimental methods and key findings in 2D ASI}

In this section, we provide a broad overview of standard experimental methods to probe both the ASI microstate (Sec.~\ref{Static}) and the resulting dynamics (Sec.~\ref{Dynamics}). We will also review key experimental findings focusing on 2D systems. 

\subsection{Static Magnetic Characterization}\label{Static}

\subsubsection{Scanning Probe Characterization}

\begin{figure}[t]
\includegraphics[width=0.4\textwidth]{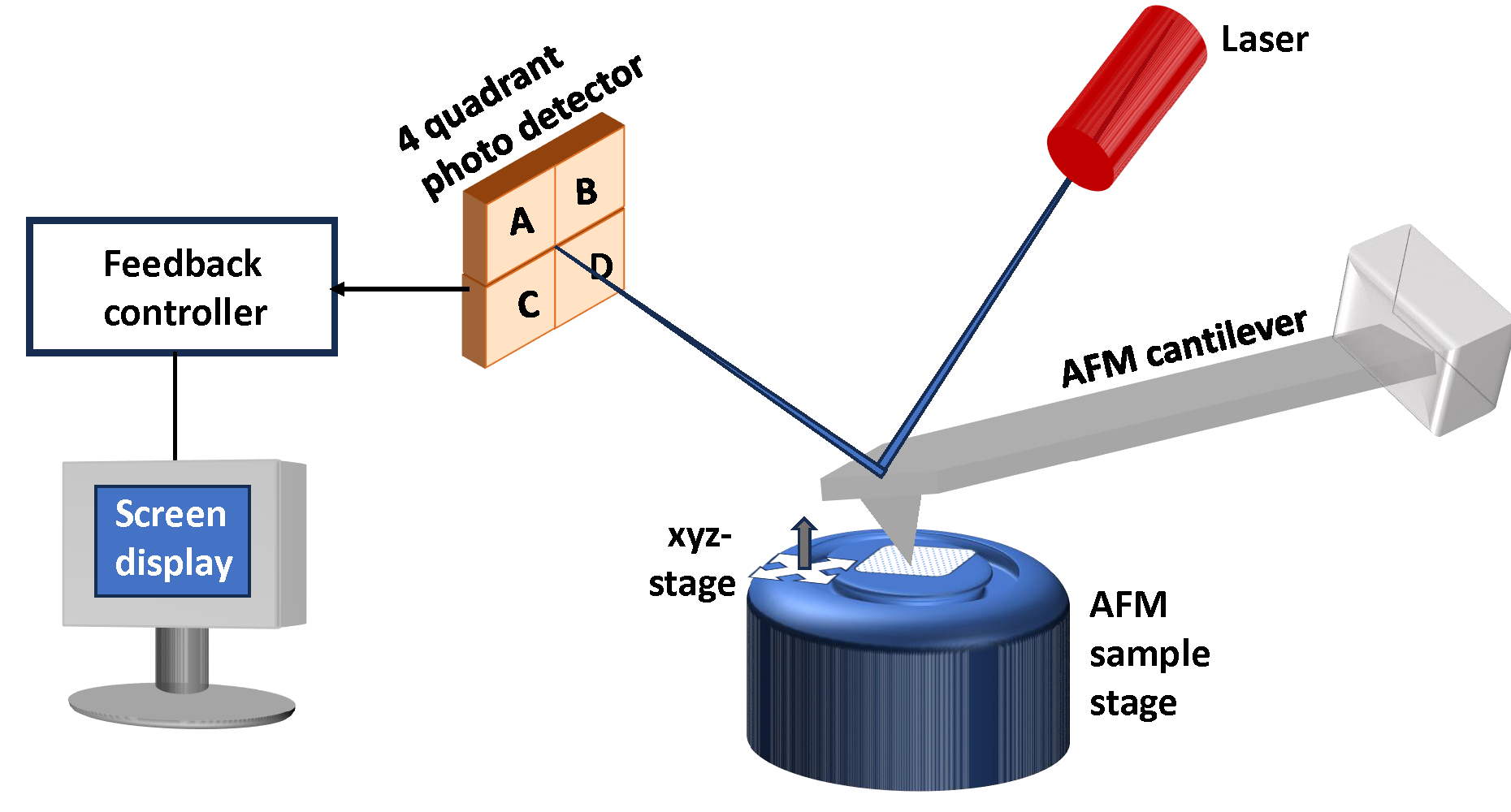}
	\centering
	\caption{{Schematic of an atomic force microscope (AFM). An AFM creates images by moving a  cantilever with a sharp tip across a sample's surface. As the tip makes contact with the surface, it bends the cantilever, altering the amount of laser light reflected detected by a photodiode. To maintain a consistent signal, the cantilever's height is adjusted, causing the measured cantilever height to follow the surface topology.}\label{AFM}}
\end{figure}

\begin{figure}[t]
	\includegraphics[width=0.5\textwidth]{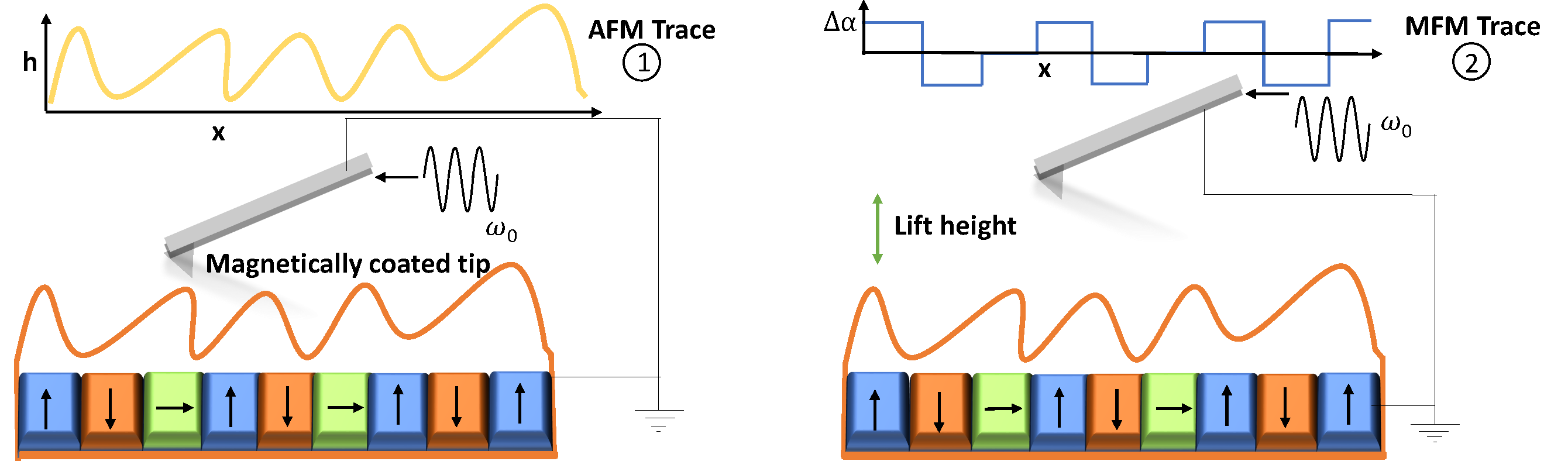}
	\centering
	\caption{{Schematic of the standard two-pass MFM mode: In the first pass (left), the topography of the sample is detected by tapping mode atomic force microscopy in which the amplitude of the cantilever's oscillation is controlled in feedback loop that varies the tip-sample distance. In the second pass (right), the cantilever is lifted away from the sample surface (lift height) in order to map the long range interactions, and the phase change of the oscillating cantilever is measured at a constant tip-sample separation.    }\label{MFM_fig}}
\end{figure}

Several methods have been developed for imaging magnetic structures with varying sensitivities on different lateral length scales.
These methods are broadly categorized into beam- and scanning probe-based techniques. Among the scanning probe-based techniques, magnetic force microscopy (MFM) is particularly notable for its extensive use in locally characterizing magnetic nanostructures and imaging the magnetic field distribution at the surface of magnetic materials\cite{Martin_1987, Saenz_1987, Belliard_1997, Freeman_2001}. This technique is based on atomic force microscopy (Fig.~\ref{AFM}) and relies on probing the long-range magnetostatic force between the magnetic sample and a magnetically coated tip, which is positioned at a constant height above the sample surface.
The MFM procedure involves two main steps as shown in Fig.~\ref{MFM_fig}. First, the surface topography is obtained using the standard atomic force microscopy (AFM) in tapping mode, which exploits van der Waals interactions between the tip and the sample (Fig.~\ref{MFM_fig} - left). {The schematic in Fig.~\ref{AFM} shows the basic principle of atomic force microscopy. A cantilever with a sharp tip approaches the surface of the sample which is placed on a $x-y$ piezo stage. A laser beam is focused on the back side of the cantilever and reflected onto a position sensitive photodetector. As the AFM tip  scans over the sample having with height variations, the deflection of the cantilever changes and, hence, the position of the laser spot on the photodetector will also move. A controlled feedback loop ensures a constant tip-sample interaction during the scanning to obtain high resolution images.} In the next step, a second scan is performed with the tip lifted away from the sample, such that van der Waals interactions are negligible, and the tip experiences only long-range magnetic and electrostatic interactions (Fig.~\ref{MFM_fig} - right). During this scan, the initial topography profile is repeated  at a constant probe–sample separation to obtain the phase change of the oscillating cantilever. One of the key advantages of MFM is its high spatial resolution (down to $\sim$ 10 nm)\cite{Schwarz_2008}, sensitivity ($\sim$ 10 pN)\cite{Seo_2005}, and its ability to operate at different temperatures and magnetic fields for studying magnetization processes\cite{Ares_2015}.

As described earlier, the long-range interactions (i.e., force gradients) are measured in MFM between the magnetic tip and the magnetic sample. These interactions are recorded in the second pass by measuring the shift in frequency ($\Delta \omega$), amplitude ($\Delta A$) or phase ($\Delta \alpha$) with respect to the initial parameters ($\omega_0$ , $A_0$ and $\alpha_0$, respectively) of the oscillating cantilever. {However, $ \Delta \alpha$ is the most commonly used representation of the magnetic contrast in the second pass of MFM.} 

When there are no interactions between the tip and the sample, the oscillating probe can be modeled as a point-mass spring. In this case, its behavior can be defined by a classic non-linear, second-order differential equation based on Newton's second law of motion\cite{Garcia_2002}. The resonant frequency of the cantilever is given by

\begin{equation}
\omega _0 = \sqrt\frac{k}{m},
\end{equation}

\noindent
where $k$ is the spring constant and $m$ is the effective mass. When the cantilever is excited sinusoidally at a frequency $\omega$, the tip oscillates sinusoidally with a corresponding amplitude $A$ and exhibits a phase shift $\alpha$ with respect to the drive signal applied to the piezoelectric actuator. The equation of motion describing the output from the cantilever sensor is given by

\begin{equation}
m\ddot z + \frac{m\omega_0}{Q}\dot z + kz = F_{ts}\ + F_0 cos(\omega t),
\label{Eq:osci}
\end{equation}

\noindent
where F$_0$ is the amplitude of the driving force, $Q$ is the quality factor of the free cantilever. F$_{ts}$ contains the tip–surface interaction forces. In the absence of tip–surface forces, F$_{ts}$ = 0, the above equation represents the forced harmonic oscillator with damping. The quality factor $Q$ is related to the damping factor $\gamma$ as 

\begin{equation}
Q = \frac{m\omega_0}{2\gamma},
\end{equation}
\noindent
where $\gamma$ introduces the influence of the environmental medium, which could be ambient air, liquid or ultrahigh vacuum (UHV). 

The steady state solution for the forced oscillator [Eq.~(\ref{Eq:osci})] is

\begin{equation}
z(t) = z_0 + A cos(\omega t + \alpha).
\end{equation}
\noindent

The dependence of the amplitude with the excitation frequency is given by

\begin{equation}
A = \frac{F_0}{m[(\omega_0^2 - \omega^2)^2 + (\frac{\omega \omega_0}{Q})^2 ]}\\
\end{equation}
and the phase {in free space} is given by\cite{Magonov_1997, Whangbo_1998}
\begin{equation}
\mathrm{tan}   \alpha_0 (\omega) = \frac{\omega \omega_0}{Q (\omega_0^2 - \omega^2)}.
\end{equation}
\noindent
The above expressions can be simplified by oscillating the probe at $\omega$ = $\omega_0$ and the phase {$\alpha_0$ } becomes
\begin{equation}
\alpha_0 (\omega_0) = \frac{\pi}{2}
\end{equation}
\noindent
The introduction of tip-sample interactions (F$_{ts}$) changes the oscillation, thereby modifying the system's response. For small displacements (z) with respect to the rest position (z$_0$) of the cantilever, the force can be written as follows after a Taylor expansion:

\begin{equation}
F_{ts}  \approx \frac{dF_{ts} (z)}{dz}|_{z=z_0} z(t).
\end{equation}
\noindent

Thus, the equation of motion [Eq.~(\ref{Eq:osci})] becomes 
\begin{equation}
F_0 \mathrm{cos}(\omega t) = m\ddot z + \frac{m\omega_0}{Q}\dot z + [(k - \frac{dF_{ts} (z)}{dz}) z(t)].
\end{equation}
\noindent

Here, a number of forces can be acting between the tip and the sample simultaneously, such as van der Waals, magnetostatic and electrostatic interactions. F$_0$ describes the amplitude of the driving force and $\frac{m\omega_0}{Q}$ is the damping factor. The phase shift in the presence of $\frac{dF_{ts}}{dz}$ is given by 

\begin{equation}
\mathrm{tan} \alpha (\omega) = \frac{m\omega \omega_0}{Q (k + \frac{dF_{ts}}{dz} - m\omega^2)}.
\end{equation}
\noindent
If the probe oscillates at $\omega_0$ and $\frac{dF_{ts}}{dz} \ll k$, Eq.~(14) can be substituted into Eq.~(22). For a given value of $k$ we find with $\frac{dF_{ts}}{dz}$, the phase to be 

\begin{equation}
\mathrm{tan}\alpha (\omega_0) = \frac{k}{Q\frac{dF_{ts}}{dz}}.
\end{equation}
\noindent

{The shift in the phase at the resonance frequency is given by }

\begin{equation}
\Delta\alpha = \alpha_0 (\omega_0) - \alpha (\omega_0)
\end{equation}

{Combining Eqs. (20) and (24) finally provides the expression for the phase shift as } 

\begin{equation}
\Delta\alpha = \frac{\pi}{2} - tan^{-1}(\frac{k}{Q\frac{dF_{ts}}{dz}}) \approx \frac{Q}{k} \frac{dF_{ts}}{dz}
\end{equation}

{The frequency-modulated modes in MFM and other scanning probe techniques utilize frequency shifts to capture magnetic contrast in MFM images\cite{Giessibl_2000}}. From Eq.~(21), the effective spring constant of the cantilever ($k_{eff}$) can be defined as

\begin{equation}
k_{eff} = k - \frac{dF_{ts}}{dz} |_{z=z_0}.
\end{equation}

The above equation leads to a softer cantilever if the force gradient is positive and a harder cantilever for the negative force gradient\cite{Garcia_2002}. The modified frequency $\omega_0^{'}$ is given by

\begin{equation}
\omega_0^{'}= \sqrt \frac{k-\frac{dF_{ts}}{dz}}{m}
\end{equation}
\noindent

Assuming $\frac{dF_{ts}}{{dz}} \ll k$, Taylor expanding Eq.~[(25)] yields:

\begin{equation}
\Delta \omega = \omega_0^{'} - \omega_0 = - \frac{\omega_0}{2k} \frac{dF_{ts}}{dz}.
\end{equation}
\noindent

{The above equation shows that the resonance frequency of a weakly perturbed harmonic oscillator is related to the gradient of the interaction. A change in the effective resonance frequency will alter the probe's oscillation amplitude according to Eq. (18). The amplitude's dependence on the  excitation and effective resonance frequency suggests a potential mechanism for elucidating how the oscillation amplitude is influenced by the strength of the interaction force, which can also be interpreted in terms of tip-sample separation.}

The most common method to excite AFM cantilevers involves oscillating the chip holding cantilever using a piezoelectric transducer. However, this method is not particularly advantageous in low-temperature systems, as instabilities may arise from the thermal contraction of mechanical parts. Several techniques have been developed to excite AFM cantilevers at low temperatures. One effective technique is photothermal excitation of the cantilever using two laser sources. In this technique, one beam is focused at the end of the cantilever for deflection measurement, while the second beam is focused near the base of the cantilever for excitation induced by the photothermal effect \cite{Ratcliff_1998,Ramos_2006}. Celik et al. recently introduced a novel pressure-based method for exciting cantilevers in dynamic AFM mode\cite{Celik_2017}. This method was demonstrated to perform well by imaging magnetic domains in Co/Pt multilayers and an Abrikosov vortex lattice in a BSCCO (2212) single crystal at 4 K. Additionally, the authors proposed a simplification to the optical design of cryogenic AFM/MFM by using a single laser beam for both radiation pressure excitation and detection of cantilever deflection in AFM imaging. For high-temperature imaging of magnetic phenomena and transitions, Peltier or resistive heaters are employed, enabling measurements to be conducted in situ over a wide temperature range from room temperature to 520 K\cite{Weller_2016}. 

In MFM experiments, it is crucial to separate electrostatic influences from the MFM signal. At typical probe-sample working distances, magnetic and electrostatic interactions can have comparable values depending on the electric and magnetic properties of the sample. An electrostatic contribution is present whenever the probe and sample exhibit different work functions, resulting in a contact potential difference (CPD). This electrostatic interaction can obscure other long- or short-range interactions\cite{Celik_2017,Kim_2017}. While applying an appropriate voltage bias between the tip and sample can compensate for CPD in a homogeneous sample, it is not applicable when the sample surface consists of more than one material, as the CPD value varies across the surface. In heterogeneous samples, it is crucial to consider this problem for correct image interpretation, particularly for low magnetic moment materials\cite{Cervenka_2009}. One of the methods involves  a combination of Kelvin probe force microscopy and MFM (KPFM/MFM), which distinguishes between electrostatic and magnetic contributions. This method records both the CPD map and the real compensated MFM image, as it cancels the electrostatic interaction between the probe and sample at every point of the image, thus obtaining a pure magnetic signal\cite{Jaafar_2011}.

In order to enhance the lateral resolution and sensitivity beyond the limit of commercial MFM probes, custom-made MFM probes have been developed. These probes allow for the modification of  stray field distribution and intensity, thereby enabling more precise  quantitative MFM studies. One of the most common approaches involves customizing the magnetic coatings, where the magnetic properties of the materials are varied\cite{Panchal_2017,Freire_2016,Nagatsu_2016,Precner_2014}. This is particularly beneficial because partially coating MFM probes or depositing multiple layers help control low/high moment states and confines the eminent stray-field primarily to the probe's apex\cite{Panchal_2017,Wren_2017}. 

\begin{figure}[t]
	\includegraphics[width=0.4\textwidth]{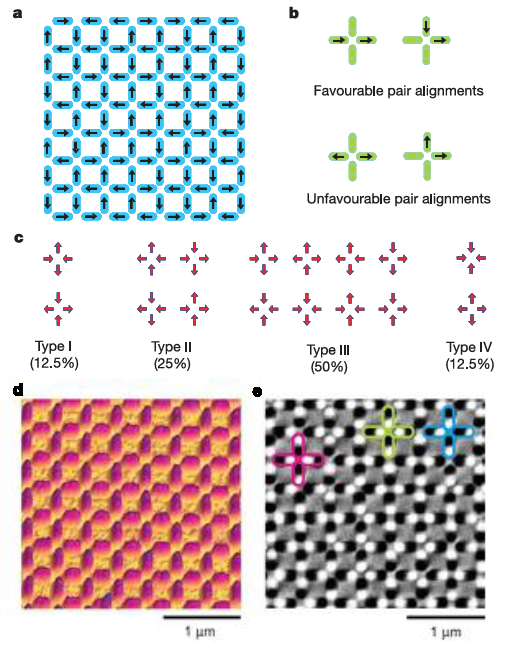}
	\centering
	\caption{{(a) Illustration of the nanomagnet configuration used to create artificial square lattice. The arrows indicate the magnetic moments revealed by the corresponding MFM image shown in (e). (b) A pair of moments on a vertex illustrating energetically favorable and unfavorable dipole interactions between the pairs. (c) Four distinct topologies for the moment configurations on a vertex of four islands with a total multiplicity of 16. (d) and (e) An AFM and the corresponding MFM image. The single-domain character of the islands is depicted by the division of each island into white and black halves, representing north and south poles.} Adapted from Ref.~[\onlinecite{Wang_2006}].}\label{MFM-results}
\end{figure}

The interpretation of MFM images is most straightforward for samples with a magnetization orientation perpendicular to the imaged surface. The applications of MFM include thin films with perpendicular magnetic anisotropy (PMA), patterned magnetic structures\cite{Albrecht_2015}, and nanodot and antidot arrays\cite{Hu_2011,Kaidatzis_2016,Rodriguez_2016, Goiriena_2017}. Among the most extensively studied magnetic patterned structures is  ASI. In particular, MFM has been used to study geometric frustration, ordering of effective magnetic charges and various collective dynamics\cite{Ladak_2010,Diaz_2001,Park_2017}; see also Sec.~\ref{microstate}. In their ground state, some of the most commonly studied structures such as square and honeycomb lattices obey the ice-rule\cite{Tanaka_2006}, but can be excited into higher energy states by external stimuli, for instance, by applying a magnetic field (see Sec.~\ref{static}). As discussed in Sec.~\ref{deterministic}, another possibility is writing and erasing of individual bits by applying in-plane field below the nanoisland saturation-field and selectively switching nanoislands with an MFM probe\cite{Wang_2016} or to introduce topological defect-driven magnetic writing\cite{Gartside_2018}.


 In square ASI (Fig.~\ref{MFM-results}), a pair of moments on a vertex can be oriented in a manner either to maximize or minimize the dipole interaction energy between them, see Fig.~\ref{MFM-results}(b). {The most energetically favorable configuration is when the moments of the pair of islands are oriented such that one is directed towards the center of the vertex and the other is oriented outwards. In contrast, it is energetically unfavorable when both moments are directed inward or both are directed outward.} For the vertex as a whole, there are four distinct topologies depending on the configuration of four moments with a total multiplicity of 16, as shown in Fig.~\ref{MFM-results}(c)\cite{Wang_2006}. MFM allows us to image the orientations of all the moments in a large area. The white and black spots in the MFM image indicate the south and north poles of the ferromagnetic islands, thereby confirming the single-domain nature of the islands. The data demonstrates that the possible vertex configurations [Fig.~\ref{MFM-results}(c)] can be directly experimentally observed in the actual system as is shown in Fig.~\ref{MFM-results}(e).

 We note that other lattice types have also been investigated extensively by MFM. For example, the Kagome lattice features three-moment coordination at each vertex and exhibits a richer phase diagram associated with effective magnetic charges intrinsic to the three-moment vertices\cite{Qi_2008, Moller_2009}. Recently,  new ASI geometries  designed to give rise to new phenomena, have been studied using MFM. For these more specialized works, we refer the reader to the literature, e.g., Refs. [\onlinecite{Gilbert_2014,Park_2017,zhang2021string,zhang2023artificial}].

\subsubsection{Lorentz Microscopy}

Lorentz transmission electron microscopy (TEM) provides the ability to image the magnetic induction with very high spatial resolution, down to 1 nm. This is suitable for mapping the structure of magnetic induction of sub-micrometer- or nanometer-scale magnets in ASI arrays\cite{Phatak_2011, Qi_2008, Li_2019, Puttock_2022}. Besides, TEM is also capable of characterizing crystal structure and chemical composition at the same time, thereby allowing for a comprehensive understanding of magnetic materials. As compared to the standard TEM, Lorentz TEM has two ways to operate: 1) the objective lens is switched off to create a low-field sample environment and a mini objective lens is used for imaging, or 2) a dedicated Lorentz TEM is designed to have a special objective lens to create the field-free environment in the vicinity of magnetic samples\cite{Phatak_2016}. 

The Lorentz TEM technique allows to obtain both qualitative and quantitative information of magnetic states. From the classical perspective, when electron beams pass through a perpendicular magnetic induction in a uniform thin magnetic film (typically film thickness $<$ 200 nm), the electrons are deflected by the Lorentz force ({Fig.~\ref{fig:lorentz}}). The Lorentz force $F_L$ that a moving electron with charge $e$ experiences is given by

\begin{equation}
F_L = evB_{\perp},
\end{equation}

\noindent
where $v$ is the velocity of the electron; $B_{\perp}$ is the component of magnetic induction perpendicular to the velocity vector of the electron. The magnitude of Lorentz deflection angle $\theta_L$  of the electron beam is given by\cite{Phatak_2016, Park_2010} 
\begin{equation}
\theta_L = \frac{e\lambda}{h}B_{\perp}t,
\end{equation}

\noindent
where $\lambda$ is the electron wavelength, $h$ is Planck's constant and $t$ is the thickness of the film. Typical Lorentz deflection angles are in the range of tens to hundreds of microradians, which is two or three orders smaller than the Bragg deflection angle for electron diffraction from crystal structures (in the range of a few milliradians)\cite{Graef_2000}.

Under an out-of-focus condition, the deflected electrons from a thin film with multiple domains converge or diverge on the image plane to yield brighter and darker contrast at the positions of domain walls, as sketched in {Fig.~\ref{fig:lorentz}(a)}. Since in the majority of studied ASIs, each nanomagnet has a single-domain state, the darker contrast is more visible at the edge of each nanomagnet in an out-of-focus Lorentz TEM image\cite{Qi_2008, Li_2019,Puttock_2022,Pollard_2011,Daunheimer_2011,Drisko_2017} [see Fig.~\ref{fig:lorentz}(b)]. The disappearance of brighter contrast is attributed to the influence of electrostatic phase arising from the boundaries of the patterned nanomagnet. This classical Lorentz deflection model offers a good understanding of magnetic imaging using the so-called Fresnel mode in conventional TEM (CTEM)\cite{Tanase_2009, Ngo_2016}. The magnetic contrast in the Fresnel image is invisible at an in-focus condition, and the locations of darker and brighter contrast are reserved in under-focus and over-focus images, as seen in Fig.~\ref{fig:lorentz}(c). The Fresnel mode can be considered a qualitative observation mode. Nevertheless, the quantitative information on the magnetic induction can be obtained through phase retrieval theory, as discussed in the following:

\begin{figure}[t]
	\includegraphics[width=0.5\textwidth]{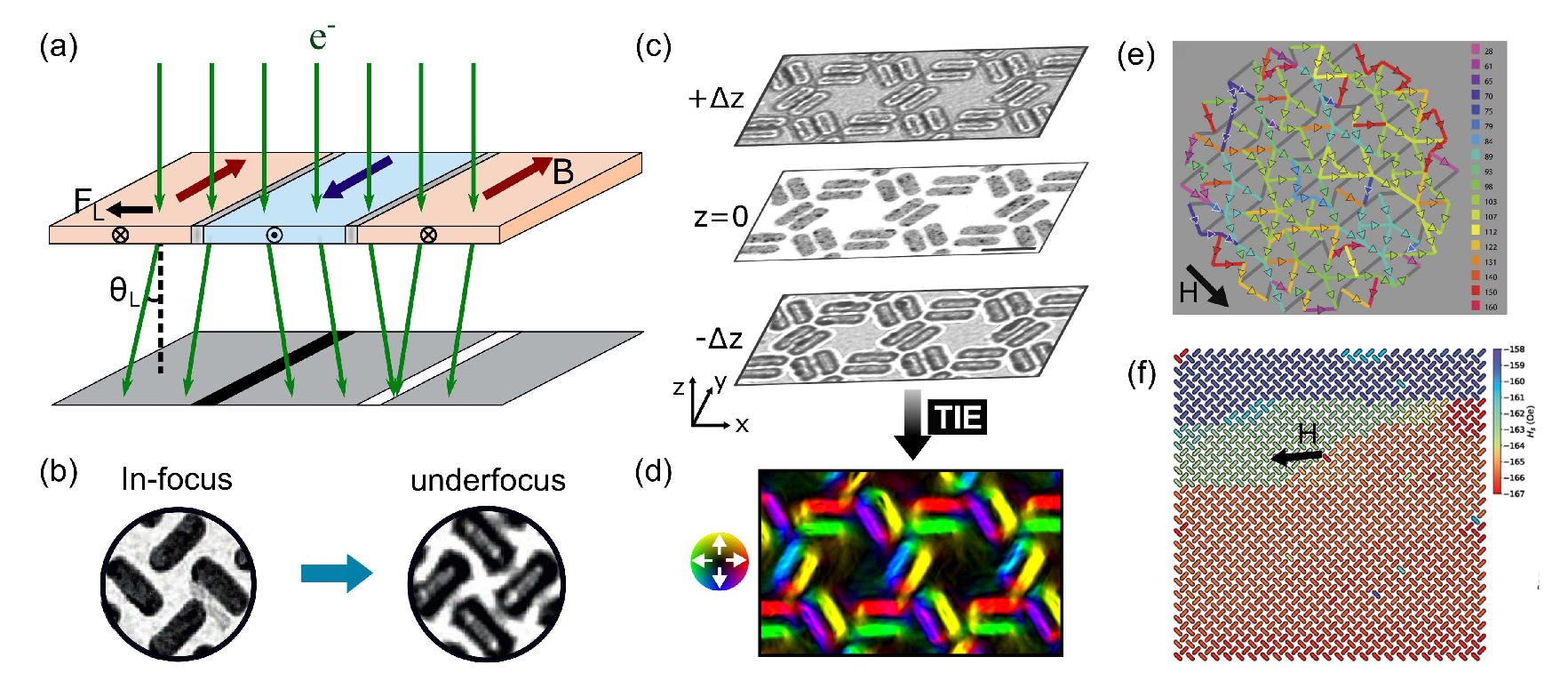}
	\centering
	\caption{(a) Schematic of Fresnel imaging mode in TEM, showing the bright and dark contrast formed at the domain walls under an out-of-focus condition. (b) The measured in-focus and under-focus TEM images of a pinwheel artificial spin ice, wherein mainly darker contrast is visible in each nanomagnet. Adapted from Ref.~[\onlinecite{Li_2019}] (c) Through-focus Lorentz TEM images of dimer Kagome artificial spin ice, including over-focus($+\Delta z$), in-focus ($z=0$) and under-focus ($-\Delta z$). (d) Retrieved magnetic induction map by the means of transport-of-intensity equation showing the local spin state of each nanomanget and stray field between nanomangets, forming a flux closure state. Adapted from Ref.~[\onlinecite{Li_2022}]. (e) Field-induced magnetization reversal in an artificial quasi-crystal showing dendritic avalanches. Adapted from Ref.~[\onlinecite{Brajuskovic_2016}]. (f) Ferromagnet-like magnetization switches driven by an varying external magnetic field in a pinwheel artificial spin ice. Adapted from Ref.~[\onlinecite{Paterson_2019}].}\label{fig:lorentz}
\end{figure}

Aharonov and Bohm in 1959 found that the phase of an electron wave was shifted upon traversing a space with electromagnetic potential based on a quantum perspective\cite{Aharonov_1959}. The total phase shift ($\varphi_t$) is composed of electrostatic ($\varphi_e$) and magnetic ($\varphi_m$) phase shifts. The electrostatic shift is a result of the mean inner potential and inhomogeneity (such as non-uniform thickness, defects, nanomagnet boundaries, etc.) of materials. The total magnetic phase can be described by\cite{Graef_2000}

\begin{equation}
\varphi_t = \varphi_e + \varphi_m = \frac{\pi}{\lambda E_t} \int_{L} V(r_\perp,z)dz - \frac{e}{\hbar}\int_{L} A(r_\perp,z)d\hat{z},
\end{equation}

\noindent
where $V$ and $A$ represent electrostatic and magnetic vector potentials, respectively. $E_t$ is the total energy of the electron beam, and integration is done along the straight line $L$ aligned with the incident beam. $r_\perp$ is a position vector which is perpendicular to the electron beam. The phase of the electron wave is associated with the intensity gradient along the trajectory under paraxial approximation, which can be satisfied in Lorentz TEM since the Lorentz deflection angle is very small, as discussed above. The relationship between the total phase shift and intensity can be expressed by using the transport-of-intensity equation (TIE)\cite{Paganin_1998, Humphrey_2014},

\begin{equation}
\nabla\dot(I_0\nabla\varphi_t)=-\frac{2\pi}{\lambda}\frac{\partial I}{\partial z},
\end{equation}

\noindent
where $I_0$ is the intensity of the Lorentz TEM image at the in-focus condition, $\frac{\partial I}{\partial z}$ represents the derivative of the intensity along the electron-beam direction (along the $z$-axis). The derivative of the intensity is evaluated by taking a through-focal series of Lorentz TEM images at under-focus, in-focus and over-focus conditions, as shown in Fig.~\ref{fig:lorentz}(c). The gradient of magnetic phase shift, $\nabla\varphi_m$, is related to the magnetic induction, which can be written as\cite{chapman_2014}
\begin{equation}
\nabla\varphi_m=-\frac{e}{h}(B_{\perp}\times \hat{n})t,
\end{equation}

\noindent
where $n$ is the unit vector aligned with the electron beam. Therefore, the in-plane magnetic induction, which is perpendicular to the electron beam, can be retrieved from the magnetic phases after removing the electrostatic phase shift, as shown in Fig.~\ref{fig:lorentz}(d). One way to remove electrostatic $\varphi_e$ involves flipping the sample over and taking a second image in the same region, in which only the magnetic contrast should reverse sign\cite{Humphrey_2014}. 

Another approach is to use off-axis electron holography. This technique uses the interference of two coherent electron beams from the sample and vacuum regions to generate a hologram, which must then be post-processed to reconstruct the magnitude and phase information\cite{Ngo_2016}.  The magnetic field of square ASI can be resolved using this technique\cite{wesels_2022}. 

The Lorentz mode in scanning TEM (STEM), such as differential phase contrast\cite{Krajnak_2016} and 4D STEM\cite{Almeida_2016} methods, can provide quantitative characterization of magnetic induction $B$ with a better spatial resolution ($\sim$1 nm) than Fresnel modes ($\sim$10 nm). However, so far, Fresnel imaging is more commonly used as the experimental operation is more straightforward ({Fig.~\ref{fig:lorentz})}.

Lorentz TEM is most commonly used for the observation of the static magnetic microstates of ASI, including the variation of local spin structures (such as domain walls and vortex defects) within each nanomagnet or magnetic bars\cite{Qi_2008, Li_2019, Puttock_2022,Brajuskovic_2021}, and interaction (stray field) between nanomagnets\cite{Phatak_2011,wesels_2022, Li_2022,Pollard_2011}. On the other hand, one of significant advancements of Lorentz TEM is the real-space and real-time observation of quasi-dynamic magnetization, such as domain wall motion and changes in spin textures, under varying external stimuli, e.g., magnetic field, temperature, electric current and field. Lorentz TEM studies have been extensively leveraged to visualize the magnetization reversal behavior that is driven by external magnetic field in ASI. It is straightforward to generate an in-plane magnetic field with respect to the sample by using a magnetizing TEM holder and out-of-plane magnetic field that is aligned with the optical axis in the microscopy by applying a current to the objective lens. The creation and annihilation of monopole charges and Dirac strings in a square ASI have been investigated, unveiling that the positive and negative monopoles are always paired and connected by Dirac strings\cite{Pollard_2011}. The aperiodicity was also found to impact the avalanches of magnetization in a connected artificial quasi-crystal lattice, forming dendritic cascades\cite{Brajuskovic_2016} as seen Fig.~\ref{fig:lorentz}(e). Furthermore, the field-driven behavior in a pinwheel ASI shows the coherent magnetization reversal which is analogous to that in natural ferromagnets at a certain field angle [Fig.~\ref{fig:lorentz}(f)]\cite{Li_2019, Paterson_2019}.

Other possible external stimuli that are compatible with Lorentz TEM are temperature and electric current/voltage. Thermal dynamics are important to understand physical mechanisms of the formation of equilibrium states in ASIs (see Sec.~\ref{static}). Although the magnetic configurations of the ASIs after a thermal quench were studied via Lorentz TEM\cite{Macauley_2020, Brajuskovic_2021}, the in-situ imaging of thermal magnetization dynamics of ASI using Lorentz TEM have not been reported yet. This research area has been extensively studied using photoemission electron microscopy with X-ray magnetic circular dichroism (PEEM-XMCD)\cite{Kapaklis_2014}, see Sec.~\ref{Dynamics}. The in-situ electrical current/voltage-driven magnetization dynamics in the ASIs is also intriguing for future potential application in spintronic devices and can be achieved by using a voltage-biased TEM holder. This research direction using Lorentz microscopy still remains to be explored in the future. One challenging issue for Lorentz TEM experiment is the sample preparation. In order to be electron-transparent, the ASIs  typically have to be patterned on a 30-100 nm silicon nitride membrane, which makes the process of electron-beam lithography more challenging (see also Sec.~\ref{Nanofab}).

While still in its infancy we want to add that there are efforts to use Lorentz TEM for studies of magnetization dynamics, which are the focus of the next section (Sec.~\ref{Dynamics}). Here, we note that real-time imaging of dynamic magnetization in ASIs using Lorentz TEM represents an exciting, challenging, and largely unexplored area of research. Magnetization dynamics span a wide temporal range, from ultrafast exchange interactions at the 10 ps scale to slower domain wall motion occurring over nanoseconds to hundreds of milliseconds.

Recent advancements in TEM techniques aim to enhance the temporal resolution of imaging. Ultrafast electron microscopy (UEM), for instance, employs stroboscopic imaging with laser-excited\cite{Aseyev_2020} or laser-free\cite{Fu_2020} electron pulses, achieving sub-picosecond temporal resolution. An advanced laser-free UEM by integrating a phototype RF electron beam pulser has been developed to record spatiotemporal LTEM images of magnon dynamics driven by a 5.26 GHz frequency RF excitation in a patterned permalloy strip\cite{liu2025correlated}, with temporal resolution down to around 10 ps. The real-space observation of nanoscale spin-wave dynamics uncovers the correlation between the oscillation of topological domain walls and the generation of spin waves. It also reveals the behavior of the spin-wave propagation, interference and reflection within the domain interior. However, UEM primarily probes reversible magnetization processes. Due to the low electron count per pulse, stroboscopic imaging requires repeated measurements and the integration of numerous pulses to achieve sufficient contrast. This technique is well-suited for studying spin-wave dynamics in ASIs within the tens of ps to tens of ns time range (see Sec.~\ref{theory}).

Electrostatic subframing systems offer another avenue for time-resolved imaging by deflecting the electron beam across different regions of the detector in rapid succession. This technique can theoretically achieve sub-microsecond resolution\cite{reed_2019}, making it ideal for investigating irreversible magnetization reversal and domain wall dynamics in nanomagnets and magnetic bars of ASI.

\begin{figure*}[t]
\includegraphics[width=\textwidth]{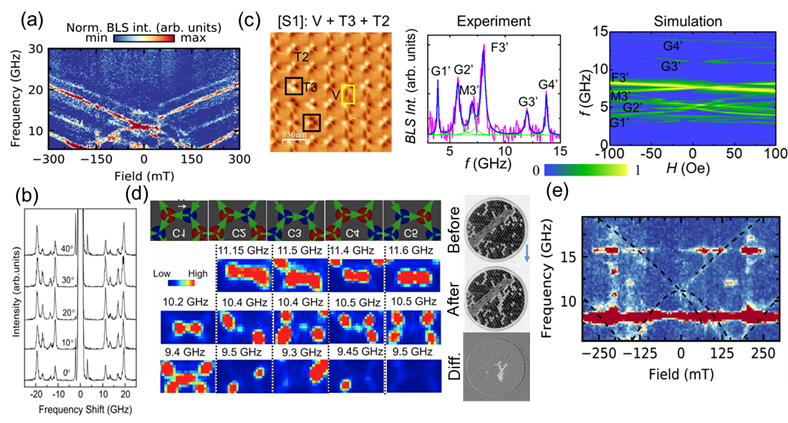}
	\centering
	\caption{(a) Thermal magnon spectrum probed by microfocused BLS. Adapted from Ref.~[\onlinecite{lendinez2023nonlinear}]. (b) Representative wavevector-resolved BLS spectra measured at an external field of 3~kOe. Adapted from Ref.~[\onlinecite{li2016brillouin}] (c) Demonstration of microstate-specific thermal magnon spectral fingerprinting in square ASI by BLS. Adapted from Ref.~[\onlinecite{mondal2024brillouin}] (d) Spatially resolved BLS intensity maps at 25 mT for fixed frequencies in different configurations (see top panel). In the color-coded images, red (blue) indicates the maximum (minimum) BLS intensity. Corresponding X-ray photoemission electron microscopy images taken at remanence, both before and after applying a microwave signal at 9.3 GHz, along with their difference, are shown on the right side of panel (d). Adapted from Ref.~[\onlinecite{bhat2020magnon}] (e) Driving microwave frequency-dependent magnon spectra in the nonlinear regime reveal nonlinear multi-magnon scattering.  Adapted from Ref.~[\onlinecite{lendinez2023nonlinear}].}\label{fig:BLS}
\end{figure*}

\subsection{Magnetization dynamics}\label{Dynamics}

\subsubsection{Brillouin light scattering}

In contrast to Raman spectroscopy, which probes higher-energy optical phonons and molecular vibrations in the THz range, Brillouin light spectroscopy (BLS) detects low-energy excitations, such as spin waves (magnons) and acoustic phonons, typically in the GHz range. BLS has become a standard method for studying magnons in various magnetic materials. The process, governed by energy and momentum conservation, involves either the creation (Stokes process) or annihilation (Anti-Stokes process) of magnons upon inelastic scattering of an incoming photon with energy $\hbar \omega_{in}$ and momentum $\hbar k_{in}$. The resulting energy $\hbar \omega_{out}$ and momentum shift $\hbar k_{out}$ of the inelastically scattered photons provides information about the probed spin wave ($\hbar \omega_{sw}, \hbar k_{sw}$): 

\begin{equation}
    \hbar\omega_{out}=  \hbar\nu_{in}\pm  \hbar\omega_{sw},\label{energy equation}
\end{equation}

\begin{equation}
    \hbar k_{out}= \hbar k_{in}\pm \hbar k_{sw}.\label{momentum equation}
\end{equation}

To analyze the small frequency shift of spin waves (typically in the GHz range), a tandem Fabry-P\'{e}rot interferometer (TFPI) is used. The TFPI consists of two highly reflective, parallel etalons. A spectrum is recorded by varying the separation between the two mirrors of each etalons. The two interferometers are arranged at an angle, allowing an effective suppression of higher order transmissions. The transmission of light through these mirrors can be described by the periodic Airy function that allows light to only pass if the mirror spacing is a multiple of half its wavelength. The absolute value of the frequency shift is determined by comparing the detected signal with the frequency of a reference signal that directly enters the interferometer.


A key advantage of BLS over microwave techniques is its exceptional sensitivity, which allows the detection of thermally activated, incoherent spin waves even in systems without external excitation\cite{sebastian2015micro}. In addition to that, BLS can be combined with a microwave source to excite the system, facilitating the detection of resonantly driven magnon modes. Furthermore, BLS is compatible with many different measurement modalities including microfocused BLS, wavevector resolution, phase resolution and time resolution. We review key features of these measurement methods in the following; for more detailed discussion, we refer to the dedicated literature, for example Refs.~[\onlinecite{Madami,Jungfleisch_BLS,sebastian2015micro}].

In microfocused BLS, a high numerical aperture objective is used to detect sub-micron dynamics by coupling the system to a 2D scanning stage and rastering the porbing laser beam over the sample\cite{sebastian2015micro}. This enables to reconstruct 2D spin-wave intensity profiles [see, for example, Fig.~\ref{fig:BLS}(d)], which can be easily compared to 2D power maps obtained by micromagnetic simulations.

Wavevector-resolved BLS is enabled by the momentum conservation, Eq.~(\ref{momentum equation}). The laser wavelength in a BLS system is fixed, hence, the wavevector of the incident light $|k_{in}|$ is constant. The wavevector resolution can be achieved by varying the direction of the incident beam with respect to the sample surface. In a thin film, the translation symmetry is broken at the sample surface, resulting in a lack of conservation for the out-of-plane component of the wavevector. However, the in-plane component of the wavevector remains conserved. Therefore, the momentum transfer can be determined by projecting the wavevector of the incident light onto the plane of the sample surface\cite{Jungfleisch_BLS}.


BLS is a phase-sensitive process, which enables the reconstruction of the phase of probed magnons. This is achieved by 
interference of two signals - a coherent reference signal with constant phase and the inelastically scattered light carrying the phase information of the probed magnons\cite{Serga_2006}.

BLS can also operate in a stroboscopic mode, allowing for the measurement of spin-wave dissipation and the propagation of short spin-wave pulses. Time resolution is achieved through a time-of-flight measurement, which tracks the elapsed time between the external excitation of a spin-wave pulse by a microwave pulse and the relative arrival time of inelastically scattered photons from the spin-wave pulse at a specific laser beam position. Due to the weak BLS scattering cross section, reconstructing the temporal evolution from a single spin-wave pulse is not feasible; instead, multiple measurements must be accumulated. To ensure consistent starting conditions for each measurement, sufficient time must elapse between successive pulses. 



\subparagraph{Brillouin light spectroscopy on ASI:}
BLS has been employed to study various aspects of dynamics in the ASI. As mentioned earlier, a key advantage of BLS is that it can be used to obtain the eigenfrequencies of ASI lattice by measuring the thermal magnon spectra. This is unlike microwave spectroscopy where a microwave antenna must be used to interrogate the dynamic response; hence, the excitation/detection of specific modes is affected by the excitation efficiency and subsequently the microwave magnetic field torque on the magnetic moments in the magnetic material (see also Sec.~\ref{microwave}). Lendinez et al. demonstrated that the thermal magnon spectrum of a square ASI measured by microfocused BLS is much richer than the corresponding spectrum obtained by microwave spectroscopy\cite{lendinezAPL2021}. Even more importantly, they showed that the BLS spectrum closely resembles the eigenmode spectrum obtained by micromagnetics. A typical thermal magnon spectrum for a permalloy square ice lattice, where the biasing field is applied along one of the principal axes, is shown in Fig.~\ref{fig:BLS}(a). The intense red lines in the color-coded image show maximum BLS intensity, which is directly proportional to magnon population and, hence, represent the eigenmodes of the square ASI lattice system \cite{lendinez2023nonlinear}.

Furthermore, thermal spectra are typically acquired in wavevector-resolved BLS measurements\cite{montoncello2023brillouin,negrello2022dynamic}. Figure~\ref{fig:BLS}(b) shows the typical intensity curve displaying both the Stokes and anti-Stokes peaks of the BLS spectrum when the external  field is applied along the direction of the horizontal elements of the Ni\textsubscript{81}Fe\textsubscript{19} square ASI islands. Four nearly flat spin-wave modes for different incident angles appear corresponding to the horizontal and vertical islands of the ASI \cite{li2016brillouin}. Due to the change of the incident angle of the laser beam and conservation of momentum, the inelastically scattered light  carries information about the wavevector of the probed magnons\cite{sebastian2015micro}. Moreover, recent experiments by Mondal et al. using thermal BLS on ASI have demonstrated the potential for spectral fingerprinting specific microstates in the spin-wave spectrum, as illustrated in Fig.~\ref{fig:BLS}(c) \cite{mondal2024brillouin}. Most importantly, this study demonstrates precise control over mode frequency shifts, microstate-specific crossings, and the number of dominant spin-wave modes. It presents a method for characterizing specific magnetic microstates in ASI by detecting BLS spin-wave peaks, similar to approaches used in X-ray diffraction.


\begin{figure}[t]
	\includegraphics[width=0.5\textwidth]{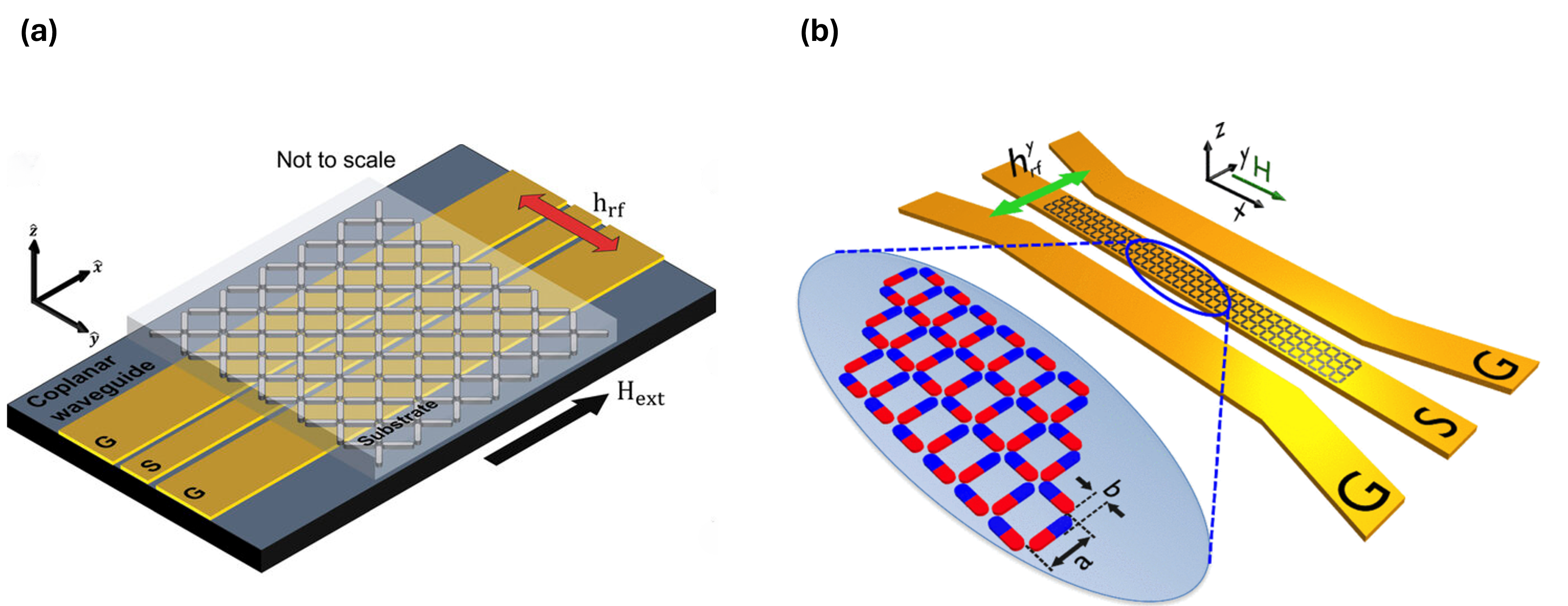}
	\centering
	\caption{Microwave spectroscopy approaches. Schematic diagram of (a) flip-chip method where a large-scale ASI is placed on top of microwave antenna. Adapted from Ref.~[\onlinecite{vanstone2022spectral}]. (b) On-chip integration of ASI by patterning the nanostructured array directly onto the signal line of a coplanar waveguide. Adapted from Ref.~[\onlinecite{jungfleisch2016dynamic}].}\label{fig:FMR}
\end{figure}


\begin{figure*}
\includegraphics[width=0.8\textwidth]{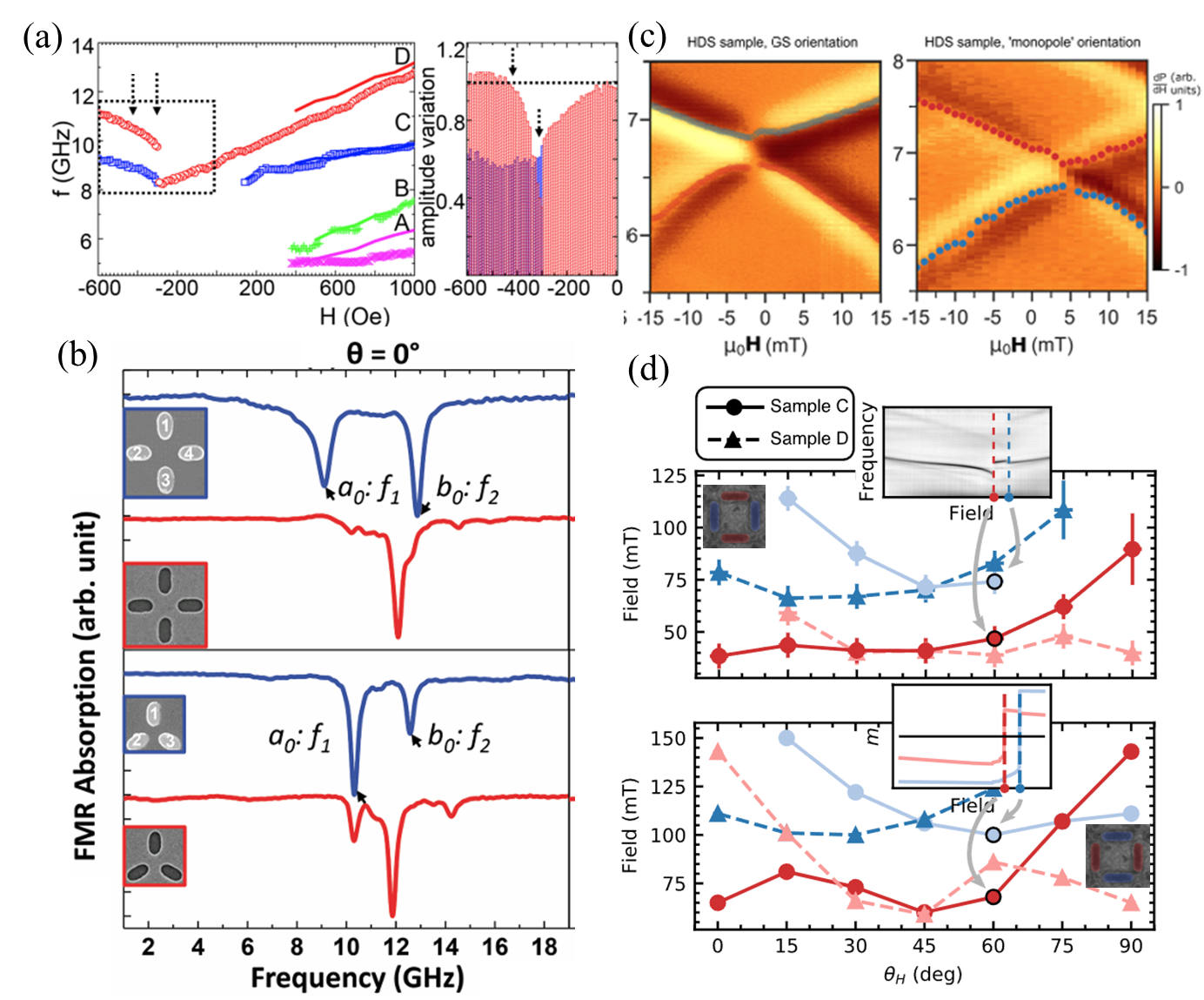}
	\centering
	\caption{(a) Field-dependent mode spectrum of a Kagome artificial spin ice obtained by VNA-FMR. The signal strength of mode C resonances for $H \leq -300$ Oe (blue columns) and mode D resonances (red columns) is extracted from the spectra within the region highlighted by the box in right panel (a). Adapted from Ref.~[\onlinecite{bhat2016magnetization}]. (b) Representative ferromagnetic resonance absorption curves of different artificial spin ice and anti-spin ice systems at saturation. Adapted from Ref.~[\onlinecite{zhou2016large}]. (c) Avoided crossing and mode-hybridization between reversed and unreversed bars in the ground state and in the monopole oriented state (characterized by four magnetic charges at each vertex) in a high-density square ASI. Adapted from Ref.~[\onlinecite{gartside2021reconfigurable}]. (d) Angular dependence of the switching fields in a bicomponent square ASI made of two different metallic ferromagnets. The red color represents nanomagnets made of Ni\textsubscript{81}Fe\textsubscript{19}, and while blue represents nanomagnets made of Co\textsubscript{90}Fe\textsubscript{10}. Adapted from Ref.~[\onlinecite{lendinez2021emergent}].}\label{VNA}
\end{figure*}

As previously mentioned, BLS is highly versatile as it can be combined with other measurement modalities, including microwave excitation (see also next section). For example, using spatially-resolved BLS, Bhat et al. showed 2D magnon mode profiles and microwave-induced avalanches in disordered Kagome artificial spin ice [Fig.~\ref{fig:BLS}(d)] \cite{bhat2020magnon}, while Kaffash et al. demonstrated angular-dependent spin dynamics in arrays of ferromagnetic nanodisks arranged on a honeycomb lattice using microfocused BLS and found that  different subsections of the lattice contribute differently to the high-frequency response of the array\cite{Kaffash_PRB_2020}.
Furthermore, BLS offers the possibility to detect a wide range of frequencies independent of the applied driving microwave frequency, unlike most microwave techniques. This property was recently used to study nonlinear magnon-magnon scattering processes and their coherence in artificial spin ice, see Fig.~\ref{fig:BLS}(e). It was shown that scattering events are determined by each nanomagnet’s mode volume and profile as well as the coupling between the dynamics of the two sublattices that exhibit dissimilar frequency-field behaviors. These result indicate that ASI can facilitate tunable directional scattering, with intriguing possibilities in reconfigurable magnonics and neumorphic computing (see also Sec.~\ref{neuro}).

\subsubsection{Microwave spectroscopy}\label{microwave}
Microwave-based techniques are standard tools to study magnetization dynamics in magentic materials including ASI. The measurement takes a shorter time compared to most optical techniques such as BLS. Several different measurement modalities can be employed for microwave spectroscopy on magnetic materials. Ferromagnetic resonance (FMR) spectroscopy is typically used to investigate magnetic nanostructures. FMR is either measured using a vector network analyzer (VNA), referred to as VNA-FMR or using a lock-in technique, where the microwave signal is provided by a microwave source and detected by a microwave diode, while the magnetic field is modulated using Hemholtz coils. The output signal is then rectified using the microwave diode detector and detected by the lock-in. Alternatively, a voltmeter can be used. VNA-FMR is usually the preferred choice to obtain a broadband spectrum; however, this technique is significantly more expensive than the simpler lock-in technique. Both methods can be used in a `flip-chip' geometry, where the ASI covers a large sample area that is placed up-side-down on a microwave antenna\cite{zhou2016large,sklenar2013broadband,bhat2016magnetization} or in an `on-chip' geometry, where the ASI patterned directly onto the antenna\cite{bang2018thickness,bang2019angular,jungfleisch2016dynamic,montoncello2018mutual}. Figure~\ref{fig:FMR} compares the two configurations. The advantage of the flip-chip geometry is that it is versatile and does not require any wirebonding or probes to connect the microwave antenna. On the flip side, the ASI network needs to cover a much larger area as compared to an on-chip integration, which makes the EBL writing process much time-consuming and, hence, more costly.

A microwave signal is then applied to the microwave antenna. This microwave signal is accompanied by a microwave magnetic field that exerts a torque on the magnetic moments in the ASI islands. The torque is maximal when the moments are perpendicular to the microwave magnetic field. If the resonance condition is fulfilled, the moments start to precess, see Eq.~(\ref{LLG}), and less microwave power is transmitted/reflected. The transmitted or reflected microwave signal is then measured by the vector network analyzer or microwave diode. By varying the bias magnetic field and excitation frequency, it is possible to experimentally access the different modes in the magnetic nanostructures. {Figure~\ref{VNA}(a)} presents a summarized broadband FMR spectra measured on a large array of interconnected 2D kagome ASI \cite{bhat2016magnetization}. All modes exhibit a monotonically increasing behavior with the magnitude of $H$, except for mode D in the range of 0 to -300 Oe. The abrupt change in frequency and slope of mode D indicates the onset of magnetization reversal in the ASI. The signal strengths of modes C and D are also depicted right panel in {Fig.~\ref{VNA}(a)}. Additionally, by engineering the geometrical arrangement of ASI or anti-ASI structures, the resonance frequencies can be tailored due to significant changes in the spin configurations\cite{zhou2016large}. A few representative FMR absorption curves for various artificial spin ice and anti-spin ice structures at saturation are shown in {Fig.~\ref{VNA}(b)}. Now an interesting question arises: are these SW modes or nanobars interacting? Gartside et al. addressed this question by analyzing different microstates in a width-modulated bi-component ASI \cite{gartside2021reconfigurable}. They demonstrated that mode hybridization between reversed and un-reversed wide bars leads to the formation of distinct upper and lower frequency $v$-shaped branches, as shown in {Fig.~\ref{VNA}(c)}, using broadband FMR in a flip-chip configuration. Furthermore, Lendinez et al. studied two dissimilar ferromagnetic metals (Ni\textsubscript{81}Fe\textsubscript{19} and Co\textsubscript{90}Fe\textsubscript{10}) arranged on complementary lattice sites in a square ASI \cite{lendinez2021emergent}. They found that the interaction between the two sublattices produced unique spectral features attributed to each sublattice. The variation of switching field values as a function of in-plane angle, presented in {Fig.~\ref{VNA}(d)}, highlights that the interaction and dynamics can be fine-tuned by appropriately selecting materials. In contrast, single-element lattices exhibit switching of both sublattices at similar field values, demonstrating the absence of such distinct interactions.

Another technique that has gained widespread popularity is the spin torque-ferromagnetic resonance (ST-FMR) technique. Here, a microwave charge current is directly passed 
through a heterostructure \cite {liu2011spin,liu2012spin} that consists of a ferromagnet, in our case an ASI, capped with a heavy metal layer with significant spin-orbit coupling. A radio-frequency charge current is then applied in the bilayer which creates an oscillating transverse spin current in the heavy metal layer due to the spin Hall effect. This spin-polarized electron current is then injected into the ferromagnet leading to the onset of the  precession of the  magnetization. When the spin precession of the ferromagnetic layer is driven into resonant oscillations, this results in a concomitant oscillation of the bilayer resistance due to the anisotropic magnetoresistance. The time-varying -resistance $R(t)$ change leads to a dc spin rectification signal $V(t)$ across the 
length of the sample from the mixing of the resistance with the microwave current $I(t)$\cite{lendinez2019magnetization}. This technique was successfully used to study the effects of collective magnetization reversals on the high frequency dynamics in connected ASI\cite{Jungfleisch_PRAppl}. Similarly, it was shown that microwave-driven dynamics in a connected ASI system can be detected by means of spin pumping and inverse spin Hall effect, enabling a simple investigation scheme of dynamics in ASI\cite{Jungfleisch_APL}. Furthermore, this work suggests that integrating connected ASI as processing and transport devices in conventional electronics is feasible. 


\subsubsection{Photoemission electron microscopy with x-ray magnetic circular dichroism}

\begin{figure}
	\includegraphics[width=0.5\textwidth]{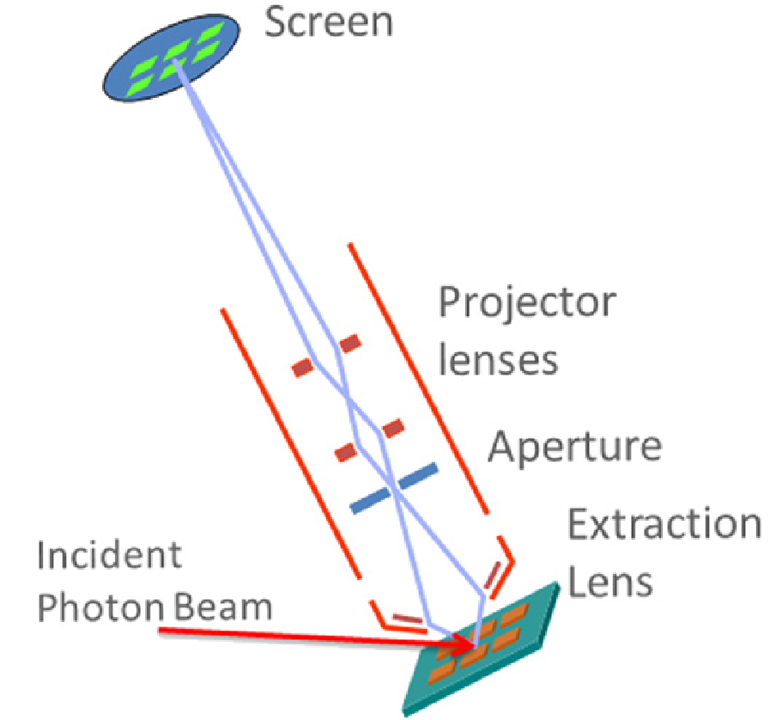}
	\centering
    	\caption{Schematic diagram of a PEEM experiment. X-rays are incident on the sample at a grazing angle, typically around 30 degrees. Electrostatic and/or magnetic lenses extract and focus the photo-emitted electrons to form a real-space image. Adapted from Ref.~[\onlinecite{cheng2012studies}]}\label{PEEM}
\end{figure}

In contrast to BLS and microwave-based techniques, photoemission electron microscopy with x-ray magnetic circular dichroism (PEEM-XMCD) provides a method to study the dynamics of ASI in real space. In this full-field imaging technique, the sample is illuminated with x-rays, and the photo-emitted electrons are collected and focused using a series of electrostatic and/or magnetic lenses\cite{scholl2002x,cheng2012studies}. This is shown schematically in Fig.~\ref{PEEM}. Typical PEEM setups can achieve a spatial resolution around tens of nanometers, which is suitable to study ASI.

For probing magnetism in 3\textit{d} transition metals like iron, manganese, and nickel (elements commonly used in ASI materials), the x-ray energy is tuned to the \textit{L} edge, which corresponds to the electronic transition where an electron is excited from the 2\textit{p} orbital to the 3\textit{d} orbital. Images are measured for incident x-rays with both right-circular (RC) and left-circular (LC) polarization, and the difference or asymmetry between these images gives the XMCD contrast. In PEEM-XMCD, the dichroic contrast $I_{XMCD}$ depends on the sample magnetization direction $\textbf{M}$ and the photon spin direction $\sigma$ as follows:

\begin{equation}
    I_{XMCD} \propto |M|\cos\varphi(\textbf{M},\sigma),\label{peem xmcd}
\end{equation}

where $\varphi$ is the angle (in the plane of the sample) between the sample magnetization and photon spin direction~\cite{scholl2002x}. In this way, the sign (positive or negative) of $I_{XMCD}$ indicates whether the in-plane magnetization direction has a component parallel or antiparallel to the direction of the incident x-rays.

\begin{figure}
	\includegraphics[width=0.5\textwidth]{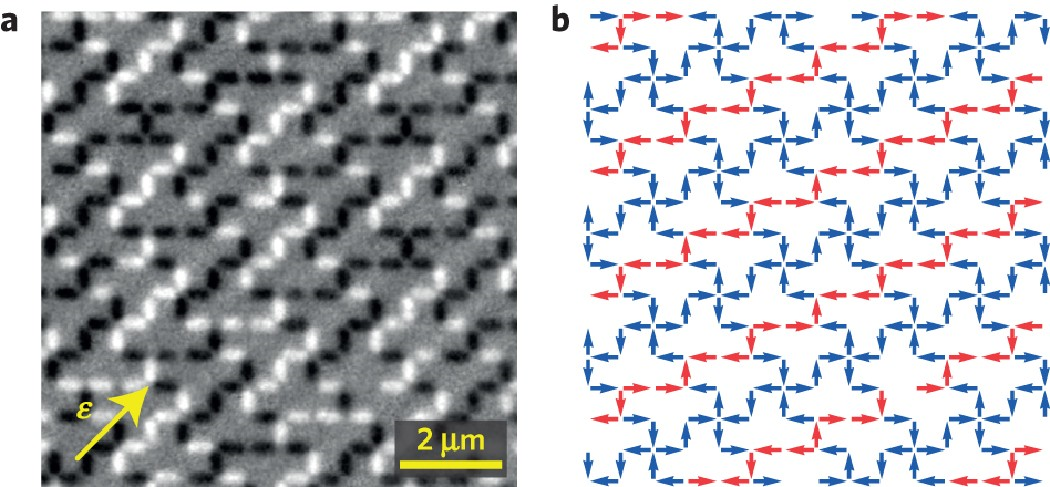}
	\centering
    	\caption{PEEM-XMCD in tetris ASI. (a) The dichroic contrast $I_{XMCD}$ is positive or negative (white or black) depending on the sample magnetization. The x-ray direction is labeled with a yellow arrow as $\varepsilon$. (b) Using the XMCD contrast and the shape anisotropy of the nanomagnets, a map of the magnetic moments can be made. Adapted from Ref.~[\onlinecite{gilbert2016emergent}]}\label{tetris PEEM}
\end{figure}

For PEEM experiments with ASI, the incident direction of the x-rays is usually chosen so that all of the segments have some component of magnetization (anti-) parallel to the beam. This allows the sign of $I_{XMCD}$ to be used along with the the shape anisotropy to determine the magnetization direction in each nanomagnet. For example, in square or tetris ASI, a common choice is to orient the beam at 45 degrees with respect to the lattice, as shown in Fig.~\ref{tetris PEEM}.

PEEM-XMCD has been used in many experiments to study both the static~\cite{mengotti2008building,gilbert2016emergent,sasaki2022formation,stromberg2024coupling} and dynamic~\cite{farhan2013direct,Kapaklis_2014,chopdekar2017nanostructured,zhang2021string,hofhuis2022real,saglam2022entropy,saccone2023real} behaviors of magnetism in ASI samples. While the references here provide a few examples of these studies, more detail can be found in Refs.~[\onlinecite{nisoli2013colloquium,lendinez2019magnetization,skjaervo2020advances,marrows2021experimental}] and the references therein.

Typical PEEM setups can incorporate both heaters and cryogenic setups, allowing the temperature to be chosen based on the ASI material under study and whether the experiment intends to measure static or dynamic behavior. Dynamic measurements are typically limited to a time resolution on the order of seconds. This is due to the time it takes to change the x-ray polarization. Faster dynamics can be achieved if measurements are taken with a single x-ray polarization, in which case the time resolution is limited by the speed of the detector and possibly the incident photon flux.

One limitation of PEEM is that it is an inherently surface-sensitive technique, because the photo-emitted electrons can only escape from the top 2-5 nm of the surface~\cite{scholl2002x,cheng2012studies}. This makes it well-suited for typical 2D ASI, but 3D systems require more advanced techniques such as x-ray magnetic laminography~\cite{pip2022x}.

\subsubsection{X-ray photon correlation spectroscopy}

Another x-ray based technique to study magnetization dynamics uses resonant x-ray scattering with a coherent x-ray beam to perform x-ray photon correlation spectroscopy (XPCS). While less common than real-space x-ray imaging techniques like PEEM-XMCD, resonant x-ray scattering from ASI can provide valuable insights into its magnetic properties. Resonant x-ray scattering is performed with the x-ray energy tuned to an elemental absorption edge. The transition metal $\textit{L}_3$ edges are common choices. Using these resonant energies enhances the magnetic sensitivity of x-ray scattering by many orders of magnitude\cite{fink2013resonant}. Resonant x-ray scattering has been used to study the magnetic configuration in square ASI with topological defects~\cite{woods2021switchable,mccarter2023antiferromagnetic}, and artificial quasi-crystals\cite{sung2018imaging}, the magnetic-field dependent switching behavior in square ASI~\cite{morgan2012magnetic,perron2013extended}, and magnetic diffuse scattering from an artificial kagome lattice ~\cite{sendetskyi2016magnetic}.

In XPCS, resonant x-ray scattering is performed with a coherent x-ray beam~\cite{madsen2020structural}. If the beam scatters off different magnetic domains, the coherence of the beam leads to interference in the scattering pattern, commonly referred to as a speckle pattern due to its speckled appearance. The fluctuations of this speckle pattern are directly related to fluctuations of the magnetic domains, so their behavior over time can be used to study magnetic dynamics in the system. Magnetic dynamics are studied by calculating the one-time intensity autocorrelation function $g_2(q,\tau)$~\cite{chen2019spontaneous,madsen2020structural,morley2017vogel}:

\begin{equation}
    g_2(q,\tau) = \frac{\langle I(q,t)I(q,t+\tau) \rangle}{\langle I(q,t) \rangle^2} = 1 + \beta |F(q,\tau)|^2,\label{g2}
\end{equation}

where $I(q,t)$ is a position in reciprocal space $q$ at time $t$. The brackets $\langle \rangle$ indicate an average over the time $t$ and over equivalent $q$ values. The function $g_2(q,\tau)$ is often written in terms of the intermediate scattering function $F(q,\tau)$~\cite{chen2019spontaneous,madsen2020structural,morley2017vogel}.

\begin{figure}
	\includegraphics[width=0.5\textwidth]{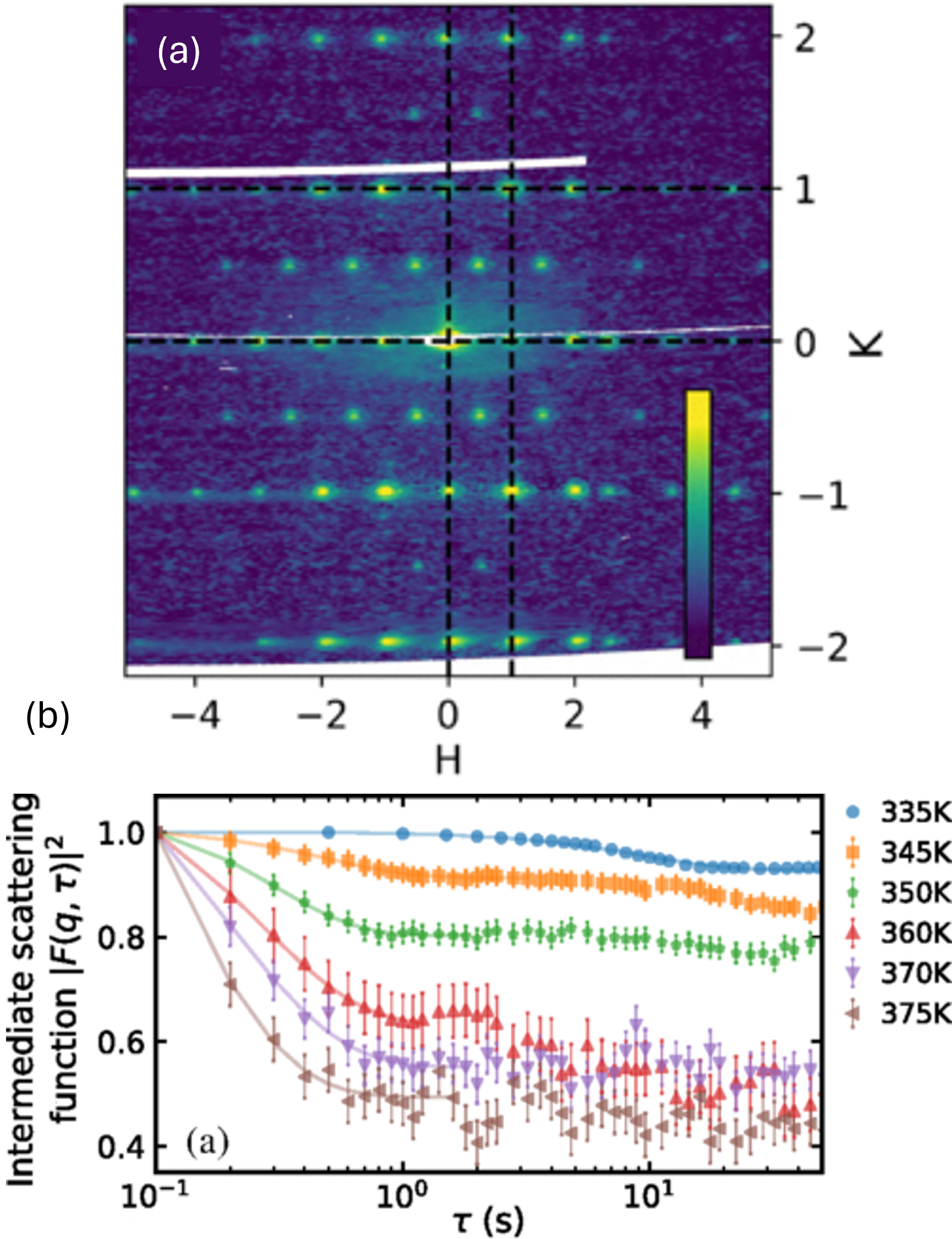}
	\centering
    	\caption{X-ray photon correlation spectroscopy from square ASI. (a) X-ray scattering pattern from square ASI. The half-integer peaks are due to the antiferromagnetic ordering in the lattice. (b) The intermediate scattering function as a function of temperature. Faster dynamics with a more rapid decorrelation time occur as the temperature is increased. Adapted from Ref.~[\onlinecite{chen2019spontaneous}]}\label{asi xpcs}
\end{figure}

XPCS has been measured to study dynamics in square ASI~\cite{chen2019spontaneous,morley2017vogel}. A typical resonant x-ray scattering pattern from a square ASI pattern is shown in Fig.~\ref{asi xpcs}(a). The ground state of the square lattice is ordered antiferromagnetically, which gives rise to magnetic diffraction peaks at half-integer reciprocal space values. To perform XPCS, a magnetic peak is selected, and the intermediate scattering function is calculated and plotted as a function of temperature, as shown in Fig.~\ref{asi xpcs}(b). At higher temperatures where dynamics are observed, the magnetic peaks have speckles~\cite{chen2019spontaneous,morley2017vogel}. The intermediate scattering function is a measure of how correlated the speckle pattern is as a function of time. For shorter time scales, $F(q,\tau) = 1$, indicating the speckle pattern has remained unchanged. For longer time scales, $F(q,\tau)$ decays to 0, which indicates the speckle pattern at later times is completely decorrelated to the initial time.

The temperature dependence of $F(q,\tau)$ can be compared to dynamic models of the ASI behavior. For example, it has been shown that some square ASI have glass-like dynamics and can be described by the Vogel-Fulcher-Tammann law~\cite{morley2017vogel}, while another study used XPCS to show the effect of superdomain-wall behavior on dynamics~\cite{chen2019spontaneous}. XPCS has the advantage of probing the collective dynamics over many domains in a system. Thus far, ASI behavior has been studied with the one-time autocorrelation function $g_2(q,\tau)$, but other methods exist such as the two-time autocorrelation function, which can be used to study systems with more complicated behavior, such as out-of-equilibrium or aging dynamics~\cite{madsen2020structural}.

\section{\label{sec:dev}Recent Advances Beyond 2D ASI: 2.5D and 3D ASI}

In recent years, ASI systems have achieved significant progress, with the introduction of novel lattice geometries such as Santa Fe, Tetris, Shakti and others (see also Sec.~\ref{intro}). These innovations have expanded our understanding of emergent magnetic monopoles and geometrical frustration in ASI. Beyond these 2D systems, research has now extended into 2.5D and 3D architectures, unlocking additional degrees of freedom for studying complex interactions, novel physical phenomena, and advanced functionalities  \cite{sahoo2021observation, Guo2023Rea, gubbiotti20242025,BhatAPL2025}. These developments are pivotal in designing systems capable of exhibiting properties unattainable in planar structures, marking an essential step toward practical applications in magnonics, data storage, and quantum computing.

\begin{figure*}
	\includegraphics[width=0.8\textwidth]{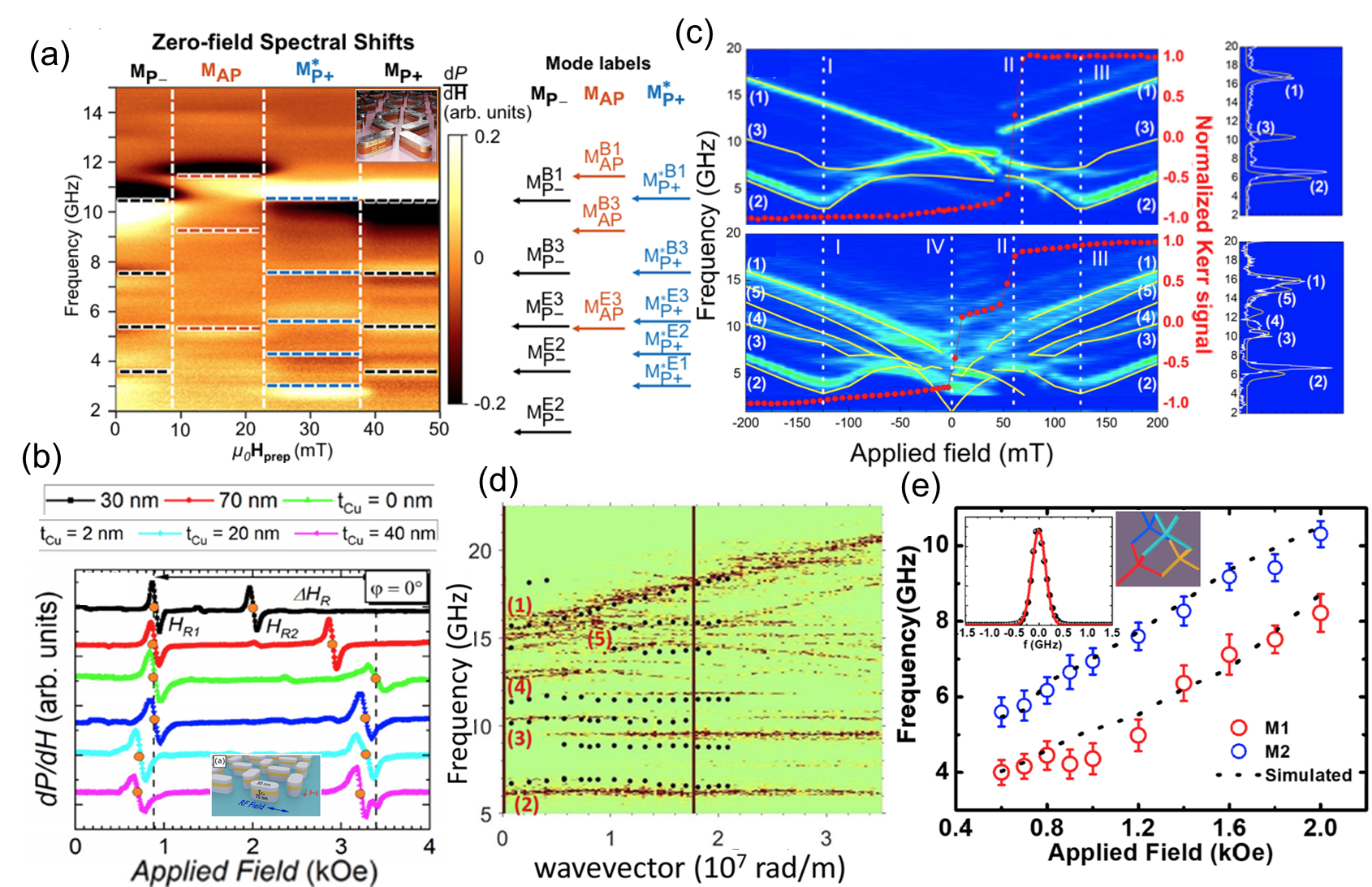}
	\centering
	\caption{(a) Remanence FMR sweep from negative to positive saturation. The $x$-axis represents the ‘preparation field’, which is applied before returning to zero field to measure the remanent spectra. The data is then modeled into multiple macrospin states and their corresponding magnetization sub-levels as detected by microwave spectroscopy. Adapted from Ref.~[\onlinecite{dion2024ultrastrong}]. (b) Representative FMR spectra at 11.5 GHz for trilayer ASI arrays with varying Cu spacer thickness ($t_{Cu}$). Adapted from Ref.~[\onlinecite{de2023tuning}]. (c) Field dependence of the BLS spectra from thermally excited SWs was measured at finite wave-vector k for both single-layer and multilayer ASI samples. Adapted from Ref.~[\onlinecite{montoncello2023brillouin}]. (d) SW frequency versus wavevector ($f$ vs $k$) dispersion for a coupled ASI-film system with a 10 nm spacer. Black dots represent experimentally measured data points, while the color shows the results of the corresponding micromagnetic simulations. The lowest lying mode shows some appreciable dispersion. Adapted from Ref.~[\onlinecite{negrello2022dynamic}]. (e) SW field dispersion for a 3D ASI system: Experimental data points are represented by symbols, with the elastic peak fitted to a Gaussian function (inset). Adapted from Ref.~[\onlinecite{sahoo2021observation}]. Schematic illustration of the samples for (a), (b), and (e) are shown the insets of the respective figures.}\label{fig:2p5D}
\end{figure*}

3D ASI systems, in particular, allow the exploration of volumetric frustration, interlayer coupling, and intricate spin textures. These multilayered systems offer enhanced coupling between layers, leading to unique magnetization dynamics and hybrid SW modes not observed in traditional 2D systems. For instance, 2.5D ASI structures, composed of multiple stacked layers, have shown improved control over magnetization behavior and SW interactions. Advanced nanofabrication methods, such as TPL, FEBID, and EBL, have enabled the creation of customized ASI structures with exceptional resolution, allowing for precise control of lattice geometries and thickness variations in fully 3D and 2.5D systems. To illustrate these advancements, {Fig.~\ref{fig:2p5D}} presents key observation of variations in spin-wave dynamics in 2.5D and 3D ASI systems, along with their tunability through various external parameters. Figure~\ref{fig:2p5D}(a) illustrates a `remanence FMR' sweep to highlight the degree of zero-field magnon reconfigurability, with all spectra taken at $H = 0$ for NiFe (30nm)/Al (35nm)/NiFe (20nm) ASI stack \cite{dion2024ultrastrong}. The $x$-axis represents the `state preparation field' ($H_\mathrm{prep}$), which is  applied prior to recording the spectra. As is obvious from the experimental results, many different states and their unique dynamics can be accessed by mcirowave spectroscopy. On the other hand, {Fig.~\ref{fig:2p5D}(b)} illustrates how the SW mode shifts as a function of the ferromagnetic layer thickness and spacer layer (Cu) in a NiFe/Cu/NiFe-based trilayer ASI device \cite{de2023tuning}. Another 2.5D system comprises of ASI coupled to an extended thin film underlayer\cite{montoncello2023brillouin}.  Figure~\ref{fig:2p5D}(c) shows the field-dependent BLS spectra of thermally excited spin waves measured at a finite $k = 4.1$ rad/$\mu$m. The results reveal reconfigurable spin-wave dispersions in a NiFe square ASI\cite{montoncello2023brillouin}. The corresponding variation in the spin-wave dispersion is shown in  {Fig.~\ref{fig:2p5D}(d)}, revealing five distinct SW modes \cite{negrello2022dynamic}.
This work  nicely illustrates how reconfigurability can be introduced by transitioning from a 2D ASI stack to a 2.5D ASI stack.
Moving to a fully 3D stack, the SW dispersion from thermal magnons in a true 3D artificial spin ice system fabricated using two-photon lithography and thermal evaporation is presented in {Fig.~\ref{fig:2p5D}(e)}. The BLS spectra reveals two dominant, nearly monotonic magnon modes which change with the external field \cite{sahoo2021observation}. The simulated mode profiles show  collective excitations throughout the complex network, while also revealing spatial quantization with varying mode quantization numbers. Detecting spin waves in a real 3D system like a 3D ASI presents two primary challenges\cite{3D_roadmap}: First, the system's 3D nature necessitates integrating multiple levels along the vertical axis, complicating the localized detection and excitation of magnons using electrical techniques.
Second, current probing methods rely on either microwave techniques or magneto-optical techniques such as Brillouin light scattering. Microwave-based methods typically measure the collective dynamics of the entire system, while magneto-optical techniques predominantly probe the topmost layers of the magnetic material. As a result, accessing the dynamics within the bulk of the structure remains experimentally challenging\cite{3D_roadmap}.

While still in their infancy, these advancements highlight the vast potential of moving beyond 2D ASI systems, emphasizing their utility in controlling magnetization dynamics, engineering reconfigurable magnonic devices, and exploring fundamental physics in new dimensions. Future research will likely integrate geometry optimization, material engineering, and external tunability to unlock further applications in magnonics, and neuromorphic computing, where ASI arrays could mimic brain-like cognitive processes. We will explore neuromorphic computing in ASI in greater detail in the next section.

$ \newline  $

\section{Neuromorphic computing in artificial spin ice}\label{neuro}

We have seen throughout this tutorial that artificial spin ice is a rich, complex system with many emergent properties, huge microstate spaces, intrinsic physical memory, GHz dynamics and compatibility with a broad range of measurement approaches. These qualities make ASI an intriguing potential candidate for providing artificial intelligence (AI)/machine learning like computation via its intrinsic physical dynamics. Leveraging complex physical systems for machine learning is the aim of a growing research effort loosely termed `physical neuromorphic computing'. In this section, we briefly introduce the concept, some background on ASI computing and describe how ASI may be used in neuromorphic computing schemes.

Recently, as the global energy cost of running machine learning/AI continues to spiral unsustainably, the search has intensified for novel computational hardware platforms which may be capable of efficiently powering AI. A large part of this effort has been devoted to identifying computational hardware which offers functionality inspired by the brain, termed `neuromorphic computing'\cite{schuman2017survey}. The brain outperforms AI at the vast majority of tasks, and consumes just 20 W instead of the MW required by GPU farms powering large neural networks. A key difference between the brain and conventional CMOS hardware is that CMOS currently stores and processes data in separate memory and processor units, with a huge portion of the computational energy budget spent on shuttling data between memory and processor rather than actually processing - the so-called `von Neumann' bottleneck. CMOS also has nowhere near the capacity for parallel processing as the brain. CMOS designers are working on developing neuromorphic-style chips using transistor technology, where memory and processor are still separate aspects of the device but many small memory caches are used and situated closer to the processor. While impressive, early results have been achieved and progress continues. However, the underlying approach and hardware are still similar to conventional CMOS and. Although the von Neumann bottleneck is reduced, it is still present. 

A parallel approach, termed `physical neuromorphic computing' aims to design and identify complex physical systems which intrinsically provide nonlinear processing and physical memory directly via their underlying physics\cite{markovic2020physics,shastri2021photonics,allwood2023perspective,finocchio2021promise,lee2023perspective}. While it is understandably at an earlier stage than CMOS transistor hardware due to its relative infancy, physical neuromorphic computing has several attractive benefits including direct integration of memory and processor, the ability to provide a far more complex and varied range of nonlinear activation functions than the simple {ReLu/sigmoid} functions used in software neural networks and the possibility of engineering intrinsic parallel coupling between processing elements through physical interactions.

\begin{figure*}
\includegraphics[width=0.9\textwidth]{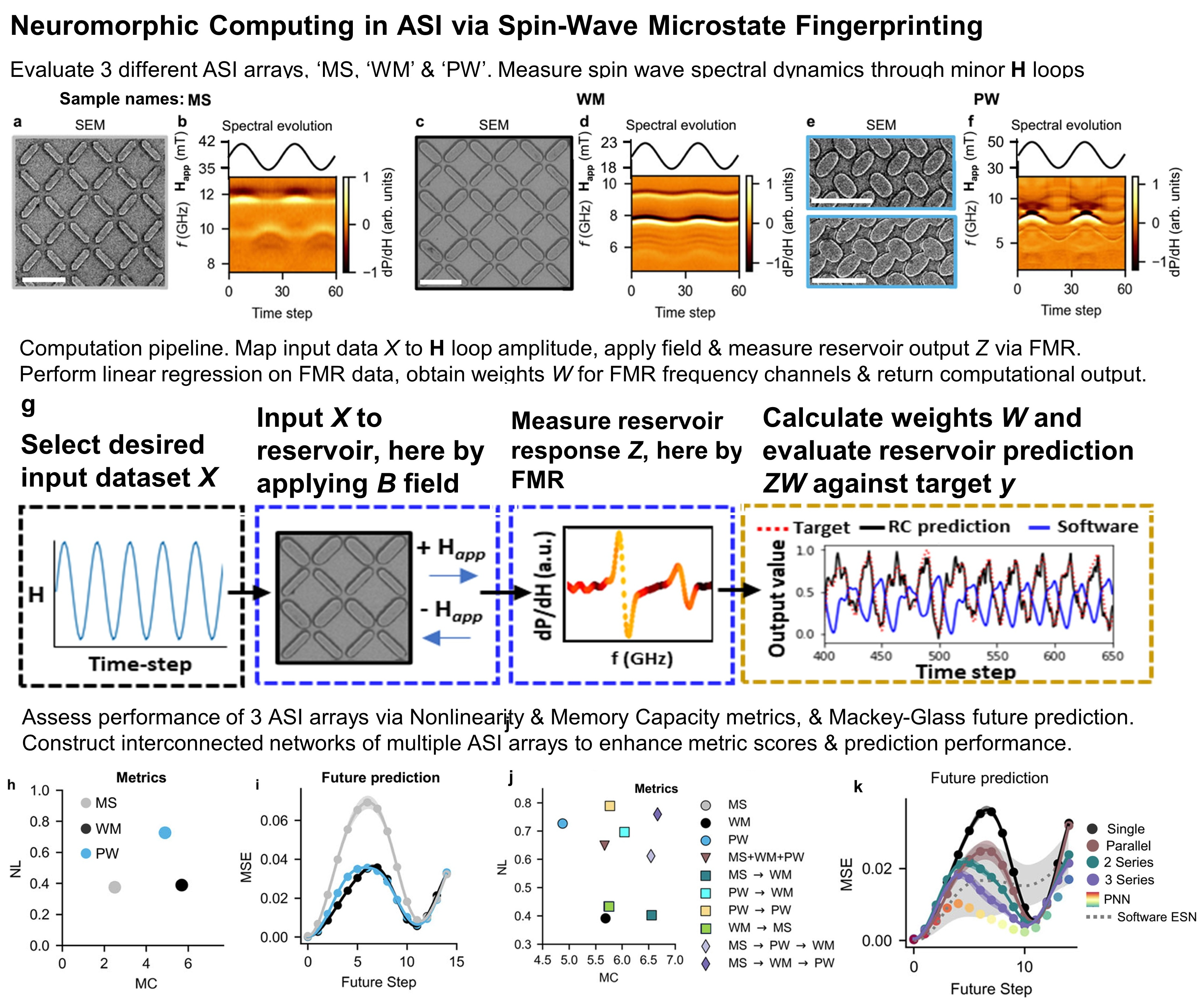}
	\centering
	\caption{Neuromorphic computing in ASI via spin-wave microstate fingerprinting. (a-f) Three ASI arrays with different microstate and magnon dynamics are fabricated (a,c,e) and their spin-wave spectra evaluated while driven through a time-series of repeated minor global magnetic field loops of varying amplitude (b,d,f). `MS': Square ASI where nanoislands only assume macrospin textures (a,b). `WM': Square bicomponent `width modified' array with nanoislands of wide and thin width. Islands are bistable between macrospin and vortex textures (c,d). `PW': Pinwheel ASI where nanoislands are bistable between macrospin and vortex textures  (e,f). (g) Pipeline schematic of the neuromorphic computing process. Input time-series data is scaled to a series of minor global magnetic field loops where the field amplitude encodes the analogue input data values. The field loops are then applied to the ASI, with FMR spectra measured after each loop to form a set of FMR spectra, one for every point in the input dataset. Linear regression is then performed on the FMR spectra and a training set of desired computational output data to obtain a single fixed weight value for each FMR frequency channel. These weights are then used to produce computational output from previously unseen test data, for instance on future prediction of chaotic time-series data such as the Mackey-Glass dataset shown here. (h,i) ASI computational performance is assessed through task-agnostic metrics (h) including nonlinearity (NL) and memory-capacity (MC), and performance on specific tasks including future prediction of Mackey-Glass chaotic time-series (i) with performance assessed by mean squared error (MSE) where lower is better. The task-agnostic metrics and task-specific MSE performance show strong differences in computational performance for different ASI arrays. (j,k) Multiple ASI arrays may be networked together, with the output of one array feeding the input of others or multiple arrays being measured in parallel. Metric performance enhances substantially for 3-layer deep ASI configurations, shown by diamonds in (j), and MSE reduces across all future prediction steps for the interconnected networks (k), with the physical neural network (PNN) where the three ASI arrays are combined in an interconnected multilayer architecture providing much stronger performance than any single array. Adapted from Ref.~[\onlinecite{stenning2024neuromorphic}].}\label{NeuromorphicFig}
\end{figure*}

Nanomagnetic arrays such as artificial spin ice are promising candidates for physical neuromorphic hardware. Nanomagnets passively retain information for 1000s of years. Neighboring nanomagnets are strongly coupled via dipolar magnetic field, enabling efficient, parallel information exchange and collective ‘in-memory’ processing at zero added energy cost - bypassing the memory-processor bottleneck. Crucially, nanomagnet arrays display collective GHz dynamics in their magnonic response, enabling information processing and transfer at rapid speeds without electron movement or Joule heating, ideal for low-energy computation and integration with existing GHz telecoms technologies. Indeed, the maths powering modern software neural networks are inspired by theoretical frameworks developed by physicists in the 1970’s to describe strongly-interacting magnetic networks\cite{sherrington1975solvable}. The early machine learning community adopted these frameworks (originally termed Hopfield networks\cite{hopfield1982neural}) and adapted and refined them towards the neural networks of today. Interestingly, Hopfield and Hinton were awarded the Nobel prize in Physics in 2024 for their ``foundational discoveries and inventions that enable machine learning with artificial neural network''. Since the early successes of machine learning, engineers have dreamt of removing the software layer of
abstraction and implementing machine learning directly in physical magnetic networks. However, until recently, the
engineering challenges of providing efficient data input/output and training schemes presented a barrier to realizing such systems.

Recent years have seen several neuromorphic computing schemes implemented in ASI and closely-related nanomagnetic arrays, using a variety of approaches to read computational output from the system including studies using the magnetic microstate\cite{jensen2018computation,hon2021numerical,yun2023electrically}, experimental schemes using the GHz magnon spectra\cite{Gliga2013,Arroo2019,stenning2024neuromorphic,manneschi2024optimising} and magneto-resistance signals\cite{hu2023distinguishing}. Nano-ring arrays which are very close to connected ASI have been studied with magneto-resistance readout\cite{vidamour2023reconfigurable,manneschi2024optimising,venkat2024exploring}. These studies reveal the capacity for ASI to provide promising neuromorphic computation performance, with the intrinsic physical memory allowing prediction of chaotic time-series, nonlinear waveform transformation and classification tasks. It is also worth noting that ASI has been studied as a platform for logic-based computation\cite{gypens2022thermoplasmonic,arava2019engineering}.

\begin{figure}[t]
	\includegraphics[width=0.5\textwidth]{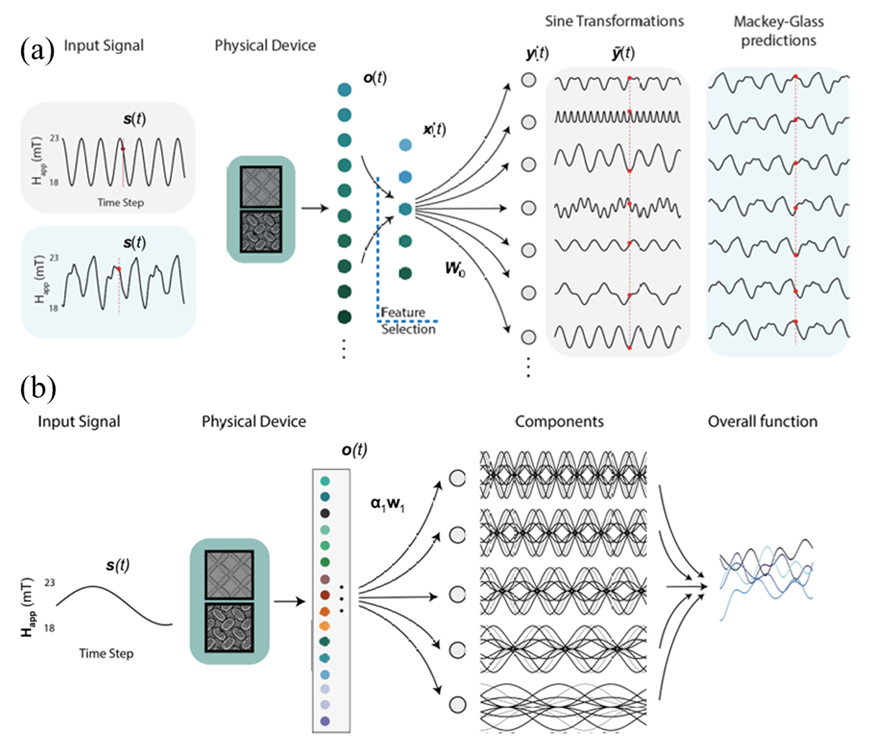}
	\centering
	\caption{Schematic of the range of distinct nonlinear, hysteretic transformations produced by the different FMR channel outputs of ASI arrays from a single time-series input. (a) shows two example datasets input to the ASI system, a sine wave (pink) and the chaotic oscillatory Mackey-Glass time-series (blue). These time series are input as minor field loops, with FMR spectra measured at each time step. Corresponding outputs of seven different targets are shown on the right side of the figure, highlighting the ability of ASI to produce many distinct complex nonlinear transformations in parallel. This ability is at the core of the computational performance of ASI when measured using spin wave spectra. (b) Examples of how the different nonlinear transformations of each time series may be combined to reproduce complex arbitrary output functions in a manner similar to a Fourier series - with the weighted sum of multiple sine waves replaced by a weighted sum of the different nonlinear responses produced by different FMR frequency channels. Adapted from Ref.~[\onlinecite{stenning2024neuromorphic}].}\label{Figure_9}
\end{figure}

In this tutorial, we will examine how one may go about designing and implementing a neuromorphic computing scheme based on ASI. 

Figure~\ref{NeuromorphicFig} shows a general schematic outline of neuromorphic computing ASI where data is input by global magnetic field and read-out via FMR spectrum, with the performance of different ASI arrays compared. The input and output schemes may be chosen freely, and the global field/FMR example here is just one way to implement computing in ASI. There are many demonstrated and proposed neuromorphic computing architectures, and for the sake of brevity as well as ease of description and implementation we will concentrate on a neuromorphic approach termed `reservoir computing'\cite{jaeger2001echo,tanaka2019recent}. A reservoir computing architecture comprises three layers, an input layer with one or more input channels where data may be introduced to the reservoir (for physical reservoirs this involves exciting the system, such as applying magnetic field, current or voltage), the reservoir itself, which acts as a set of interconnected nodes with nonlinear and recurrent connection between nodes, and an output layer with one or more channels (for physical reservoir computing these are the measurement output channels, such as electrical magneto-resistance current paths or frequency channels within an FMR spectrum). 

The reservoir must possess two key properties, nonlinearity and `fading memory'. Nonlinearity describes the reservoirs ability to perform a nonlinear transformation on the input data, which cannot be described by a simple $y = mx + c$ relationship. It is this ability that allows the reservoir computer to solve nonlinear tasks using simple linear regression methods, as the reservoir nonlinearly transforms and projects input data into a higher-dimensional output space where computational data that was not linearly separable/solvable in the input space becomes linearly separable in the output space of the reservoir. Fading memory or the `echo state' property\cite{jaeger2001echo,tanaka2019recent} describes the ability of the reservoir to provide an output response at a given timestep $t$ which contains information on previous inputs at timesteps $t-n$ - with the `fading' aspect meaning that the reservoir has a strong response to recent inputs, weaker response to inputs further in the past and eventually no response to inputs from far in the past. The fading aspect of the memory means the memory capacity of the reservoir is not simply saturated by data from very far in the past and dynamically `forgets' old data in order to be able to respond to more pertinent inputs from the recent past. If the system has no memory, it may still perform some useful tasks such as classification but it shouldn't be termed a reservoir computer - other terminology for memory-free processors include `extreme learning machines'\cite{wang2022review}.

{Reservoir computing is attractive as it does not require direct reconfigurability of the physical system during training, instead only training a separate single linear layer of weights, typically via linear or logistic regression.} This makes physical implementation easier as it is challenging to engineer reliable, accurate means for meaningfully reconfiguring the state and dynamics of a physical system, though it typically comes at a cost of reduced computational power relative to fully-trained neural networks. ASI based schemes using methods to physically write and reconfigure network `weights' through means such as electrical, surface probe\cite{wang2016rewritable,gartside2018realization}, magnonic\cite{bhat2020magnon} or all-optical magnetic switching\cite{stenning2023low,gypens2022thermoplasmonic,arava2019engineering} (see also Sec.~\ref{microstate}). 

An area where reservoirs have an advantage is in time-domain tasks such as prediction. Training larger neural networks for such history-sensitive tasks typically requires highly computationally intensive processes such as backpropagation through time. As reservoirs possess internal memory and recurrency, they allow history-dependent time-domain tasks to be accomplished through cheap regression methods, appealing for applications where reducing training cost is important.

In order to design a `good' reservoir, one wants to maximize and control key task-agnostic performance metrics. These include nonlinearity and memory-capacity as mentioned above\cite{love2023spatial}, with higher nonlinearity essential for solving classification and nonlinear transform tasks, and memory-capacity key for future prediction tasks. Typically, some combination of both properties are required, with an ideal relative ratio between the two for different specific tasks: {A challenge of designing reservoir systems is that memory and nonlinearity are somewhat inversely defined, a system has a finite signal-to-noise headroom for its output space and both the memory and nonlinearity of a system must share this finite headroom. If most of the output signal represents a nonlinear transform of the current input step, little headroom is left to reflect information based on prior time steps, e.g., a strongly nonlinear system typically has reduced memory and vice-versa.} This is termed the `memory/nonlinearity trade-off'\cite{inubushi2017reservoir} and can place limitations. There are routes to surpassing it, including improving signal-to-noise; so more headroom is available, and building interconnected networks of multiple reservoirs with different individual memory/nonlinearity scores, which can combine to give an overall network performance greater than the sum of its parts\cite{manneschi2021exploiting,gallicchio2017deep}. This networked physical reservoir approach has recently been employed in ASI systems with good results\cite{stenning2024neuromorphic}. An example of experimentally assessed memory-capacity and nonlinearity scores of ASI arrays is given in Fig.~\ref{NeuromorphicFig}, with \ref{NeuromorphicFig} (h) showing memory and nonlinearity metric scores for 3 different single ASI arrays, and \ref{NeuromorphicFig} (j) showing scores where the same 3 arrays are now interconnected to form a physical network of multiple ASI arrays where the output of one array feeds the inputs of others, leading to significantly enhanced metric scores and task performance.  

How might one increase the memory or nonlinearity of their ASI system? Increased memory can be found by tuning the ASI such that its microstates change more gradually in response to input or increasing the number of potential states. This has been achieved via a combination of the array geometry and input technique, using a pinwheel ASI and `clocked' rotating magnetic field input\cite{jensen2024clocked}, and by tuning the magnetic energetics of ASI nanoislands such that they are multi-stable beyond Ising macrospin states\cite{gartside2022reconfigurable}. The addition of metastable magnetic vortex states was introduced to form an `artificial spin-vortex ice', where vortices would nucleate in the system at a gradual rate providing longer-term memory than a purely macrospin system\cite{gartside2022reconfigurable}.

Nonlinearity may also be enhanced by array design, with disordered arrays fabricated that have a gradual gradient of nanoisland geometry and dimensions across a large array\cite{stenning2024neuromorphic}. This has the result of broadening and enhancing the range of nonlinear transforms provided by the system, with different regions of the array providing distinct physical nonlinearities -- increasing the overall nonlinearity of the reservoir\cite{stenning2024neuromorphic}. Nonlinearity may also be increased by including different distinct elements/states in the ASI array. The vortex states introduced above have a very different magnon spectra than the more linear Kittel modes of macrospin states. Including both vortices and macrospins combines both of these output responses in the GHz spectral readout and enhances system nonlinearity\cite{gartside2022reconfigurable,stenning2024neuromorphic}. This also helps by accessing different frequency-domain output channels of the magnon spectral readout, which both helps improve effective signal-to-noise by increasing the range of output frequency channels containing useful computational information and also allows the different nonlinear responses to be separated out to discrete output channels for better training performance. Essentially, engineering a more diverse set of system states and responses to input can improve nonlinearity, and engineering the system to exhibit gradual responses to input rather than abrupt switching and broadening the microstate space can improve memory capacity.

Another key property of a reservoir is its capacity to handle higher dimensional data, both on the input and output side. Reservoirs ideally want to `project' each input channel to multiple output channels with different nonlinear and hysteretic transforms on each channel, such that the training process has access to a broader range of physical responses to draw on. The capacity to accept multi-dimensional input greatly expands the range of computational tasks that the system is suited for, e.g., if one is processing an image with $n\times n$ pixels it is ideal if the system can accept $n\times n$ concurrent input channels. So far,  neuromorphic ASI systems have demonstrated the ability to produce good high-dimensional output, largely through use of GHz magnon spectra recorded through ferromagnetic resonance or spin-torque ferromagnetic resonance\cite{Jungfleisch_PRAppl}. Each discrete frequency channel of the spectra may be treated as an independent output channel with its own weight applied during training, with a few GHz-wide spectra easily providing several hundred output channels. An example of this high-dimensional output behavior, and how the different nonlinear and hysteretic responses of different FMR channels may be combined in a manner similar to the weighted sums of a Fourier series to produce arbitrary complex output functions is shown in Fig.~\ref{Figure_9}.

Electrical magneto-resistance output approaches have the ability to provide high-dimensional output, but this is challenging to realize experimentally at the lab scale as it requires the patterning of many separate current paths and electrical contacts. The electrical ASI schemes demonstrated so far typically use from one to a handful of outputs and require further development, but have the benefit that these outputs are fast and simple to measure relative to GHz spectra by microwave spectroscopy, and may be taken from different regions of the array and provide spatially-resolved readout\cite{vidamour2023reconfigurable,hu2023distinguishing}. Simulated studies which take full knowledge of the magnetic microstate as their readout can have very high dimensionality, particularly for larger arrays. However, it is challenging to extract this information experimentally -- imaging techniques such as MFM, LTEM or X-ray magnetic circular dichroism photoemission electron microscopy are slow and require expensive, cumbersome hardware (see also Sec.~\ref{static}). ASI arrays of electrically-connected islands, perhaps with multilayer magnetic-tunnel-junction-style design can be imagined to provide full microstate readout, but engineering this is challenging.

Data input is more of an unsolved issue than data output in neuromorphic ASI systems. Existing schemes typically use global magnetic fields to input data. This is an obvious choice as most labs have access to such equipment and the ASI arrays respond well to field, but it is slow, one-dimensional and hard to scale to device technologies. ASI needs better input solutions to be integrated with computing schemes in order to progress. Some exciting methods exist and now require integration with computational architectures, including electrical switching, microwave switching\cite{bhat2020magnon} and all-optical switching\cite{stenning2023low}. Improving data input is the largest open challenge in advancing neuromorphic computing in ASI and related strong-coupled nanomagnetic arrays.

Once we have designed our ASI array and identified an experimental data input and output method, it is time to train the reservoir and evaluate performance. Schematic illustration of the process is provided in Fig.~\ref{NeuromorphicFig}, where the computational performance of multiple ASI arrays with different microstate and spin-wave dynamics are assessed. The core steps to training and evaluating reservoir performance are as follows, with:

\begin{enumerate}
    \item Select a time-series or discrete data set to be the input data $X$ for your task and select a target computational output result for your training process (e.g., the result of a nonlinear transform, or correct classes for image recognition), and scale your input data to an appropriate range of the physical input value (e.g., voltage, current, magnetic field) for the reservoir you are measuring. As an example of what $X$ and $Y$ could represent, $X$ could be a set of photographs of cats and dogs, and $Y$ is a set of `labels' reading either `cat' or `dog' -- or for a time series prediction task, $X$ could be a chaotically oscillating time series and $Y$ could be a copy of that time series projected $n$ steps into the future to give a target for accurate future prediction.
    \item Sequentially input the desired dataset $X$, recording the reservoir state after each input to obtain a set of reservoir outputs $Z$ (sometimes referred to as the `reservoir states').
    \item Split the reservoir responses $Z$ and the desired computational output (often termed the target function), $Y$, into train, $(Z_{train}, Y_{train})$ and test $(Z_{test}, Y_{test})$ datasets. It is crucial that the training process is never inadvertently exposed to the test data, as doing so invalidates the testing performance and will result in artificially strong performance.
    \item With the train dataset, optimize a set of fixed weights $W$ which perform a multiply and accumulate operation on the output states. This can be achieved via a number of methods including regression (typically linear or ridge regression for continuous data such as predicting analogue time-series, or logistic regression for classifying discrete data classes such as images or spoken words) and gradient descent. These can be performed relatively simply with various software packages with some popular choices in the python libraries $sklearn$ and $PyTorch$, or hand-coded for more flexibility. The regression processes function by performing a simple linear least-squares fit to identify a weight matrix $W$ for each reservoir output channel such that the error between the desired computational output $Y_{train}$ (target data) and actual reservoir output multiplied by the weight matrix $Z_{train}W$ is minimized, e.g. $Y_{train} - Z_{train}W$ is as close to zero as possible. This is a simple single matrix solve, hence much faster and computationally cheaper than techniques like backpropagation used to train larger neural networks.
    \item Test the performance of the reservoir by applying the weights obtained during training to the test dataset, comparing the reservoir output in response to the test data $Z_{test}$ multiplied by the previously trained weight matrix $W$ to the target $Y_{test}$, e.g. evaluating $Z_{test}W - Y_{test}$. Performance is typically evaluated using mean squared error (MSE) or normalized mean squared error (NMSE) for continuous regression/prediction tasks or accuracy for classification tasks, with the error being the difference between the reservoir output and the desired target computational response $Y_{train}$.
\end{enumerate}

There are a few subtle things one must consider when testing a reservoir. The length of the training set ($n$) should be larger than the number of reservoir outputs ($m$), $n > 1.5m$ is a good starting point. If $n \leq m$, the regression will overfit the training dataset, leading to an ill-chosen set of weights which do not give good performance in the test set. A sign of this is when the train MSE is much lower than the test MSE, and one can plot train/test MSE as a function of training set length and number of reservoir outputs to better evaluate performance and overfitting. One should also ensure that the test set is reasonably large to ensure that the region of data that is evaluated in terms of performance is representative of the entire task. For example, if the input time-series is a repeated sine-wave with 30 points per period, then the test set should be at least 30 points, but ideally much longer if experimental constraints permit this.

There are a few ways the dedicated researcher can improve performance at this stage, with hyperparameter optimization and feature selection two good options. {Hyperparameter optimization involves tuning various parameters which are not part of the reservoir states, for example, input scaling factor or regression penalty term. Hyperparameters can also be related to the physical system itself, for instance if there is some voltage bias effect which may tune the behavior of the ASI array. This can be completed via a grid-search or a more sophisticated optimization algorithm.}

Feature selection involves removing certain reservoir outputs during training, in order to find a better weight solution. Quite often, two or more reservoir outputs will be highly correlated. During training, it is better to remove highly correlated channels as they do not add any additional information, only noise. Another case is where a measurement channel is only noise, in which case it is better to remove this channel prior to training. Feature selection can be achieved manually, via thresholding to only include channels with high amplitude, or removing correlated features\cite{stenning2024neuromorphic,manneschi2021sparce}.

When assessing the benefits of optimization, it is important to split the measured dataset $X$ into train, validate, and test datasets. The validate set acts like a second test set, to confirm that optimized parameters are generalizing well for the desired task rather than overfitting to a fixed single test set. It provides a second dataset which is unseen during training, and if performance is similarly strong on both the validate and test datasets, there is good likelihood that the system has generalized well rather than overfit to an arbitrary test set. For each hyperparameter/set of features, the weights are obtained using the train dataset and the computational performance of that set is obtained on the validate dataset. This is repeated for all hyperparameters. The performance of sets are compared and the best method is chosen. At this point, one performs a final evaluation of performance using the test set and this is the final performance of the model. This is necessary as it is possible to overfit the hyperparameters to the specific dataset used in the train and validate sets, which leads to bad performance on unseen data. One can go a step further and perform cross-validation, where the portion of data used for train, validate, and test is changed multiple times and an average performance over all test sets is stated. This ensures that the reservoir performs well over all parts of the input, not just the specific splitting chosen.

\section{\label{sec:sum}Outlook and summary}

The purpose of this tutorial article was to introduce readers to magnonics with artificial spin ice, which has emerged as a rapidly developing field that has received significant interest because of its basic and technological importance \cite{Kaffash_PRA,gliga2020dynamics, sklenar2019dynamics}. ASI systems, which range from two-dimensional to three-dimensional structures, offer a diverse platform for investigating emergent phenomena such as geometrical frustration, magnetic monopoles, and phase transitions. The use of sophisticated lithographic techniques to fabricate ASI arrays, as well as simulation tools that solve the Landau-Lifshitz-Gilbert equation, have allowed for a better understanding of microstate preparation, control, and microstate-driven magnetization dynamics. Our detailed description of the physical principles underlying the dynamics of artificial spin ice and its potential applications -- in reconfigurable magnonics, where magnons act as information carriers, and in neuromorphic computing, where ASI arrays can simulate brain-inspired processes \cite{gartside2022reconfigurable, stenning2024neuromorphic,jensen2018computation,hon2021numerical,hu2023distinguishing} -- aims to help researchers new to the field grasp these opportunities.


Future research in artificial spin ice aims to push the field’s boundaries even further. By integrating geometry optimization and material engineering, researchers can fine-tune lattice interactions, leading to unique magnetization dynamics. In addition to ordered geometric arrays, disordered or `glassy' artficial spin systems\cite{saccone2022direct} (or `ASG') hold interest and promise both for fundamental physics and statistical mechanics, and also for functional applications such as hardware Hopfield networks and Ising machines well-suited for challenging optimisation problems with real-world relevance such as Travelling Salesman\cite{kirkpatrick1985configuration}. The software models such as Hopfield networks currently used on these problems today are directly inspired by spin glass physics\cite{hopfield1982neural}.

The use of low-damping materials is expected to enhance spin-wave propagation, improving the performance of ASI-based devices in room-temperature applications. Additionally, hybrid systems that integrate ASI with nanoscale technologies -- including superconductors, photonics, and quantum systems -- hold promise for the development of multifunctional devices with unprecedented capabilities \cite{pal2024using, li2020hybrid, yuan2022quantum}.

Emerging dynamic control techniques, such as surface acoustic waves, spin-orbit coupling, and electric fields, are anticipated to enable real-time reconfiguration of ASI systems, unlocking the full potential of ASI-based processing and computing.

Looking ahead, research will likely focus on multilayer (2.5D) and fully 3D ASI structures, facilitating more complex interconnections and functionalities\cite{gubbiotti20242025}. The integration of van der Waals magnets and the potential for twist-engineering artificial spin ice present exciting opportunities to explore novel physical phenomena and realize new applications. Utilizing machine learning techniques is expected to accelerate the optimization of ASI designs and their scalability for practical applications. These advancements could drive the development of ASI-based systems for logic devices, data storage, quantum computing, and energy-efficient on-chip processing -- bridging fundamental physics with cutting-edge technology\cite{pal2024using, li2020hybrid, yuan2022quantum}.

\section{Data availability statement}
Data sharing is not applicable to this article as no new data were created or analyzed in this study.

\section{Author declarations}
The authors have no conflicts to disclose.

\section{Acknowledgment}
This work was supported by the U.S. Department of Energy, Office of Science, Office of Basic Energy Sciences under Award Number DE-SC-0024346. MBJ and AKM were partially supported by the National Science Foundation under
Grant No. 2339475. YL was supported by the US Department of Energy,
Office of Science, Office of Basic Energy Sciences, Materials Science and Engineering Division.

\section{References}
\bibliography{apssamp}

\end{document}